\documentclass[floats,aps]{revtex4}
\usepackage{epsfig}

\def\8{\infty}

\def\d{\partial}

\def\undertext#1{\vtop{\hbox{#1}\kern 1pt \hrule}}

\def\VEV#1{\left\langle\,#1\,\right\rangle}

\def\dd#1{\frac{d}{d#1}}

\def\be{\begin{equation}}
\def\ee{\end{equation}}
\def\bea{\begin{eqnarray} & &}
\def\eea{\end{eqnarray}}

\def\rf#1{(\ref{#1})}

\def\rf#1{(\ref{#1})}

\def\cD{{\emph D}}

\def\rfs#1{Eq.~\rf{#1}}

\begin{document}

\title{Conformal Field Theory at central charge %\lowercase{$c$}{}$=0$,\\[1mm]
{\large$\lowercase{c}=0$}\\[1mm]
and Two-Dimensional Critical Systems\\[1mm] with Quenched Disorder
%Conformal Field Theories of Critical Systems with Quenched Disorder
%\footnote{\uppercase{T}his work is supported by etc, etc.}
}

\author{V. Gurarie%\footnote{\uppercase{W}ork partially
%supported by grant 2-4570.5 of the \uppercase{S}wiss
%\uppercase{N}ational \uppercase{S}cience \uppercase{F}oundation.}
}

\address{Department of Physics, CB390 \\
University of Colorado   \\
Boulder CO 80309\\
%E-mail: victor.gurarie@colorado.edu
}

\author{A.~W.~W. Ludwig}

\address{Department of Physics \\
University of California \\
Santa Barbara CA  93106
%E-mail: ludwig@ludwig.physics.ucsb.edu
}


\begin{abstract}
We examine two-dimensional conformal field theories (CFTs)
at central charge $c=0$.  These arise typically
in the description of critical systems with quenched disorder,
but also in other contexts including
dilute self-avoiding polymers and percolation.
We show that such CFTs must in general possess, in addition to their
stress energy tensor $T(z)$, an extra field
whose holomorphic part,  $t(z)$, has  conformal weight two.
The singular part of the Operator Product Expansion (OPE) between
$T(z)$ and $t(z)$ is uniquely fixed up to a single
number $b$, defining a new `anomaly'
which is a characteristic of any  $c=0$ CFT,
and which  may be used to distinguish between different such CFTs.
The extra field $t(z)$ is not primary (unless $b=0$),
and is  a so-called `logarithmic operator' except in special
cases which include affine (Ka\v{c}--Moody) Lie-super current
algebras.  The number $b$
controls the question of whether
Virasoro null-vectors arising at
certain conformal weights
contained in the $c=0$ Ka\v{c} table may be set to zero or not, in these
nonunitary theories.  This has, in the familiar manner, implications on
the existence of differential equations
satisfied by  conformal blocks involving primary operators with  Ka\v{c}-table dimensions.
It is shown that $c=0$ theories where $t(z)$ is logarithmic, contain, besides $T$ and $t$,
additional fields with conformal weight two.  If the latter are  a fermionic pair, the
OPEs between the holomorphic parts of all these conformal weight-two operators
are automatically covariant under a global $U(1|1)$ supersymmetry.
A full extension of the Virasoro algebra by the
Laurent modes of these extra conformal weight-two fields, including  $t(z)$,
remains an interesting question for future work.

\end{abstract}

\maketitle
\newpage

%%%%%%%%%%%%%%%%%%%%%%%%%%%%%%%%%%%%%%%%%%%%%%%%%%%%%%%%%%%%%%%%%%%%%%%%
% You may put here the table of contents, just uncomment 3 lines below %
%%%%%%%%%%%%%%%%%%%%%%%%%%%%%%%%%%%%%%%%%%%%%%%%%%%%%%%%%%%%%%%%%%%%%%%%
\vspace{0.5cm}
\tableofcontents
\newpage
%%%%%%%%%%%%%%%%%%%%%%%%%%%%%%%%%%%%%%

\section{Introduction}

In the last four decades remarkable progress has been made in
our understanding  of second order phase transitions. Beginning
 with
 the scaling hypothesis put forward in
 the sixties and continuing with the subsequent  development of
the renormalization group methods,
 many questions regarding the properties of matter in
the vicinity of critical points have been thoroughly answered.  A
further milestone was set in 1984 with the development of
conformal field theory (CFT) by A. A. Belavin, A. M. Polyakov and A.~B.~%
Zamolodchikov \cite{Belavin1984,DiFrancesco}.
Indeed, the methods of CFT
provided access, in a completely nonperturbative manner, to  a vast
variety of problems involving second order phase transitions
of classical statistical mechanics in two dimensions.
The various exact solutions can be classified,
and in part distinguished, by a parameter called central charge
$c$. Physically the central charge measures the response of a
scale invariant (or critical) field theory in two dimensions to
 a change
of the geometry of
the space on which it lives.
Equivalently,  the central charge is universally related to the
coefficient of the length dependence of
 the ground state (or Casimir) energy of a critical
$(1+1)$-dimensional relativistic
 field theory living on a space of
finite length $L$ with periodic boundary conditions, i.e. where
space-time is a cylinder.  (These results are due to W. J.
Bl\"ote,  J. L. Cardy and M. P. Nightingale \cite{Cardy1986} and
simultaneously I. Affleck \cite{Affleck1986}.)

Once the central charge $c$ of  a given critical physical system
is known, the techniques of CFT make it possible, in many
important cases,  to calculate exactly all its correlation
functions in a quite straightforward
way\cite{Belavin1984,DiFrancesco}.
At the same time, finding the central charge of a given system
may not be entirely  obvious.
In some cases the central charge can be obtained by elementary
computation. This is the case for example for the Ising model,
whose central charge $c=1/2$ can be obtained using the free
Majorana fermion representation. In other cases, the central
charge of  a  system can be found from symmetry analyses, as is
the case e.g. for Wess-Zumino-Witten (WZW)
models\cite{KnizhnikZamol,DiFrancesco}, which
have found many applications in Condensed Matter Physics,
including e.g.  one-dimensional Quantum Spin Chains\cite{1DSpinChains},
 the Kondo effect\cite{Kondo},
topological Quantum Computation\cite{FreedmanEtAl},
and many others.  If not known analytically, the central charge
of a given system can also be determined numerically, using the
finite size scaling methods mentioned above.

However, there exists a class of problems where knowing the
central charge tells us close to nothing about the solution of the
CFT.
These are problems where second order phase
transitions happen (in two dimensions)
in the presence of quenched disorder.
Critical field theories describing such problems can be shown to
typically  have vanishing central charge, $c=0$. It turns out,
unlike in the case of their pure (i.e. disorderless) counterparts which have $c
\not =0$, that the knowledge of their central charge $c=0$ does not contribute
much to the solution of these theories.  Indeed, there is a
large variety of CFTs with vanishing central charge,  each of
which corresponds to a different critical point.
And with very few exceptions little is known about these theories.

These theories arise for example in the description of disordered
electronic systems.\footnote{\,In the absence of electron-electron
interactions, or when these are irrelevant in the renormalization
group sense.\\[-3mm]~} Consider a quantum  mechanical particle moving  in
a random potential
 in $d$ dimensions.
The system is described by a Hamiltonian \be
\label{DefHandHzero} H = H_0 + V(x), \  \ \ H_0 =
-\frac{\hbar^2}{2 m} \nabla^2,
\ee
where $x$ denotes $d$-dimensional space, and $V(x)$ is a  random,
time-independent potential. When describing universal critical
properties, the latter may often  be taken, without loss of
generality, to have a probability distribution which is
 a short-ranged Gaussian with zero mean
\be
\VEV{V}=0\,, \qquad \VEV{V(x) V(y)} = \lambda ~\delta(x-y)\,.
\ee
where $\left\langle ... \right\rangle$ denotes the average over
all configurations of the disorder potential $V(x)$.

All relevant information concerning
 the motion of a quantum particle can be extracted from its
 (advanced or retarded)
 Green's functions\,\footnote{\,We use  round brackets for Dirac's
bra and ket symbols, to distinguish them from the averaging
 symbols ``$\langle$" and ``$\rangle$".\\[-3mm]~}
\be \label{GreensFunction} G^{\pm}(E)(x,y) = \left( x \left|
\frac{1}{ E-H_0-V\pm i \epsilon} \right| y \right), \qquad
(\epsilon
>0)\,. \ee
These can be calculated with the help of the
following Gaussian functional integral, involving a
complex\,\footnote{\,We use the notation ${\bar \phi}(x) :=\phi^*(x)$
for the complex scalar field.}
 scalar field
$\phi(x),$ ${\bar \phi}(x)$
\begin{eqnarray}
&& (\pm i) G^\pm (E)(x,y) =
\frac{1}{Z}
\int
\cD \bar\phi {\cD} \phi~
\phi(x) {\bar \phi}(y)
\nonumber\\[1mm]
&&
\times \exp \left[  \pm i \int d^d x~
\bar \phi \left( E-H_0-V\pm i \epsilon \right) \phi \right]\,,
\label{eq:green}
\end{eqnarray}
where $d$ is the dimensionality of space, and
$Z$ is the partition function
$$
Z= {\int
\cD \bar \phi
{\cD} \phi~
~\exp \left[  \pm i \int d^d x~ \bar
\phi \left( E-H_0-V\pm i \epsilon \right) \phi \right]}.
$$
The plus or minus sign in the exponential is chosen to insure
convergence. Although this maps the problem of a Green's function
of a quantum particle moving in a random potential into a correlation
function of a field theory, this field
theory is intractable, as written.
 Indeed, the correlation function has to be computed for any arbitrary
random function $V(x)$, and the corresponding field theory is not
even translationally invariant. If, however, we concentrate on
computing Green's functions  which are averaged over all disorder
configurations $V(x)$, further progress is possible. It is not
practical, of course, to average \rf{eq:green} over random $V(x)$
directly, because of the factor $1/Z$ in (\ref{eq:green}), where
the `partition function' $Z$ is  itself a random variable.
Instead, one can employ one of the following two `tricks',
commonly referred to as `{\it replica}'-  and `{\it supersymmetry
tricks}'. In the present paper we concentrate mostly on the
supersymmetry trick, which involves rewriting the denominator of
\rf{eq:green} as a functional integral over anticommuting {
(Grassmann) } variables $\psi$, $\bar \psi$
\begin{equation}
\label{eq:part}
\frac{1}{ Z} =
 {\int
\cD \bar \psi
{\cD} \psi~
~\exp \left[ \pm i \int d^d x~ \bar \psi \left( E-H_0-V\pm i
\epsilon \right) \psi \right]}.
\end{equation}
 This brings the Green's function into the form
\begin{equation}
\label{eq:fir}
(\pm i) \  G^\pm(E)(x,y)
= \int
\cD \bar \phi {\cD} \phi
\cD \bar\psi {\cD} \psi~
\phi(x) {\bar \phi}(y)
~e^{-S_V},
\end{equation}
where the action $S_V$ is given by
\begin{equation}
\label{eq:action}
S_V =   \mp i \int d^d x~ \left\{ \bar \phi \left( E-H_0-V\pm i
\epsilon \right) \phi + \bar \psi (E-H_0-V\pm i \epsilon ) \psi
\right\}.
\end{equation}
The total partition function is unity, because the fermionic and
bosonic contributions to it cancel. Now averaging over the random
potential becomes possible with the help of { the  standard
Gaussian identity }
$$ \VEV{\exp \left[i \int d^d x~V(x) J(x)
\right]}=\exp\left[ - \frac{\lambda}{2} \int d^d x~J^2(x) \right]
$$
 valid  for an arbitrary function $J(x)$.  This yields
\begin{equation}
\label{eq:effect}
(\pm i) \ \VEV{G^\pm(E)(x,y)} =
 \int
\cD \bar\phi {\cD} \phi
\cD \bar\psi {\cD} \psi~
\phi(x) {\bar \phi}(y)
~e^{-S},
\end{equation}
where the action $S$ is given by
\begin{eqnarray}
\nonumber
\hspace{-5mm}
S&=&\mp i\!\! \int\!\! d^d x\left\{\bar \phi \left(
E\!-\!H_0\pm i \epsilon \right) \phi\! +\!\bar \psi (E\!-\!H_0\pm i \epsilon
) \psi  %\right.
%\nonumber\\[3mm]
%&\pm& \left.
\pm i \,\frac{\lambda}{ 2} \left(\bar \phi \phi \! +\! \bar \psi
\psi \right)^2 \!\right\},~~
\\[1mm]
&&(\lambda >0)\, .
%\nonumber
\label{eq:actionsusy}
%\label{actionsusy}
\end{eqnarray}
In summary, the problem of computing an averaged Green's function
of a quantum mechanical particle in a random potential in $d$
dimensional space, has been mapped into a problem of computing a
correlation function in a $d$-dimensional field theory of
interacting bosonic and fermionic degrees of freedom.  As we
already mentioned, \rf{eq:effect} is often referred to as the
supersymmetry (SUSY) approach to disordered
systems.\footnote{\,%
It should be emphasized that the action $S$ does not possess
space-time SUSY (where the translation operator is
a suitable square of the supercharge),
the SUSY that is usually understood in  high energy physics.
Rather,  it has an isotopic SUSY which involves rotating
bosonic and fermionic fields $\phi$ and $\psi$ into each other
(see e.g. \cite{EfetovBook}), and is often referred to as
supergroup symmetry. }

Theories of this kind have been extensively studied in the
literature\cite{EfetovBook}, using a variety of techniques in
various dimensionalities. Progress in accessing critical
properties can sometimes be made if a small parameter is
available, such as for example in the $d=2+\epsilon$ expansion.
The topic of this paper is two-dimensional physics, and here a
small parameter is typically unavailable.  Accordingly, one needs
to rely on nonperturbative techniques, and
CFT is expected to provide such tools. To be specific, let us
discuss in a little more detail a specific
disordered two-dimensional electronic system known to
possess a critical point.
Consider a quantum mechanical  particle moving in a plane
(coordinate $x$) in the presence of a perpendicular  constant
magnetic field
and in a random potential $V(x)$. In order to  write down an
effective field theory for this problem,  we proceed as above,
but now choosing  $H_0$ to be the Hamiltonian for a  free particle
in $d=2$ dimensions, moving in a constant magnetic field
$$
H_0 = -\frac{\hbar^2}{2m} \sum_j \left( \frac{\d}{\d x_j} + \sum_k
\frac{i \epsilon_{jk} x_k}{2 l^2}  \right)^2,
$$
where $l$ is the magnetic length.

It also turns out that the disorder averaged Green's function
$\VEV{G^\pm(E)}$,  as in Eq.~\rf{eq:effect},
does not exhibit any critical behavior whatsoever,
in this, and in the other problems discussed above.
It decays exponentially on distances larger than the
particle's  mean free path.
However,  the average of the advanced/retarded product
\be
\label{GadvGret}
\VEV{G^+ \left(E \right) G^- \left(E\right)}
\ee
can be critical\cite{KaneStone}.  Such a product can also be cast
into the form of a correlation function in a  field theory
if only  one chooses two independent functional integral
representations such as \rf{eq:fir} for the
two Green's functions
involved in the product, subject to the same disorder potential,
and then averages over disorder $V(x)$.
The resulting  field theory is  similar to \rf{eq:effect} but contains
two copies of each of the two  basic fields $\phi$ (bosonic) and $\psi$ (fermionic).
As the parameter $E$ (energy) is adjusted, the resulting field theory
goes through a critical point, called the Integer Quantum Hall
plateau transition.  This transition is experimentally
observed\,\footnote{\,Even though long-range Coulomb interactions between
the  electrons appear to modify the transition, unless they are screened
by hand, in which case they are known to leave  the non-interacting universality class unaffected
(see e.g. Ref.'s\cite{LeeWangElectronInteractions,HPWeirecent}).\\[-3mm]~}
in the Integer Quantum Hall  Effect (see e.g. \cite{HuckesteinRMP,HPWeirecent}).
Even  though much is known from numerical work\cite{IQHEnumerical}
about the critical
properties\,\footnote{\,For example, the correlation length exponent
is known to be numerically close to $\nu=7/3$.\\[-3mm]~} of this transition
(in the absence of interactions), and even though a theoretical
description in terms of a nonlinear sigma model with topological
term\cite{Pruisken,Khmelnitskii}
was given a long time ago,
an analytical solution  of  the transition
has been lacking for,  by now,  about two decades.
As already mentioned, this is due to the fact that this problem lacks
a  small parameter, and a genuinely nonperturbative
approach is unavoidable; conformal field theory
is expected to provide such a nonperturbative  tool.
Nevertheless, CFT techniques have not yielded a solution to this problem, to date.
This is due to certain `technical' difficulties which CFTs,
aimed at describing disordered critical points, present.
It is because of these `difficulties'
that exact,  nonperturbative solutions
of the infinite number of constraints
imposed by the conformal symmetry group,
have not been forthcoming as readily
as  was the case in pure (i.e. nonrandom)
critical theories\cite{Belavin1984}.
Some of these difficulties are the subject of this paper.

Even though an analytical solution of the Integer Quantum Hall
plateau transition is still lacking as of today, it has been
possible, fairly recently, to find an analytical solution of the
rather similar (but not identical) problem of the so-called Spin
Quantum Hall Effect (SQHE)
plateau transition\cite{GruzbergLudwigReadSQHE}.  The resulting theory is
a supersymmetric formulation of the 2D percolation
problem.\footnote{\,Another solution of the SQHE transition, not
based on SUSY, was later found in Ref.\cite{CardySQHE}.}
Percolation and the problem of dilute self-avoiding walks,
in fact, are two of the best understood  disordered systems in two
dimensions. In spite of this, the nature of their CFT, including for example
multi-point correlation functions, is quite poorly understood\cite{ButSee}.
Both systems have central charge $c=0$, and we will describe aspects of their
CFT below. Because the self-avoiding polymer problem is
(in a formal sense) closely related to our formulation of particle localization in
(\ref{eq:actionsusy}), let us describe this now  in some detail.
The statistics of self-avoiding dilute polymer chains in $d$ dimensions
can be described by the following SUSY Landau--Ginzburg
action\,\footnote{\,Formally, one might envision this action as
arising from (\ref{eq:actionsusy}) by analytic continuation, and a
change of sign of $\lambda$, i.e. $g := (-\lambda) >0$,  but we
will not pursue this here.} due to Parisi and
Sourlas\cite{ParisiSourlas}
\begin{eqnarray}
S &=&
\int d^d x~
\left\{\bar \phi\left(
%E-H_0
H_0-E
%\pm i \epsilon
 \right)
\phi
+
\bar \psi
\left (
%E-H_0
H_0 - E
%\pm i \epsilon
\right ) \psi + \frac{g}{ 2} \left(\bar \phi \phi  + \bar \psi
\psi \right)^2 \right\},
\nonumber\\[1mm]
(g &>& 0)
\label{eq:actionsusyParisiSourlas}
\end{eqnarray}
with $H_0$ as in (\ref{DefHandHzero}) with $\hbar=m=1$. Note that,
in contrast to the problem of the motion of a quantum particle,
described by the theory  (\ref{eq:actionsusy}), the `convergence
factors' $\pm i \epsilon$ have disappeared, and it turns out that
the analog of the `single-particle Green's function',
\begin{equation}
\label{eq:effectParisiSourlas}
\VEV{G^\pm(E)(x,y)} =
 \int
\cD \bar\phi {\cD} \phi
\cD \bar\psi {\cD} \psi~
\phi(x) {\bar \phi}(y)
~e^{-S},
\end{equation}
now exhibits  critical behavior
%is now critical
(has power-law decay, at $E=0$) and
characterizes the statistics of a polymer chain
with end points fixed at positions $x$ and $y$.

As was already mentioned, conformal field theories describing disordered
critical points in 2D typically have central charge $c=0$.
Indeed, as emphasized below (\ref{eq:action}), these theories are
constructed in such a way that their partition function is always
exactly equal to unity,
$$
 \int
\cD \bar\phi {\cD} \phi
\cD \bar\psi {\cD} \psi~
~e^{-S} =1,
$$
as a consequence of exact cancellation of bosonic and fermionic
integrals. The free energy is therefore exactly zero. This is also
true when the theory is defined on a cylinder.  Hence the central
charge vanishes.

Let us end our introductory remarks by briefly mentioning the
so-called replica approach to disordered
systems\cite{Replica}.  This involves introducing, before taking
the average over disorder realizations,  several copies of,
say,  the commuting field $\phi_\alpha$,
($\alpha =1, ..., n$),  instead of introducing the anticommuting field $\psi$,
and then taking the number $n$ of copies
to zero ($n\to 0$) (`Bosonic Replicas').
(An equivalent formulation can be obtained by using $n$ copies
%only
of the anticommuting (Grassmann) field, and no commuting fields
(`Fermionic Replicas').)
For example, introducing Bosonic replicas in (\ref{eq:green}),
which describes the Green's function of a particle moving in a random
potential, and performing the average over disorder, one easily finds that
the following functional
integral can be used as an alternative to \rf{eq:effect},
\begin{equation}
\label{eq:effectrepl}
(\pm i)  \VEV{G^\pm(E)(x,y)}
= \lim_{n \rightarrow 0}
\int \prod_{\alpha=1}^n \left [
\cD \bar\phi_\alpha
{\cD} \phi_\alpha\,
\right]
\phi_1(x) {\bar \phi}_1(y)
~e^{-S_r(n)},
\end{equation}
where the `replicated action' $S_r(n)$ is given by
\be
 S_r(n) = \! \mp i \!\int\! d^d x \left[
\sum_{\alpha=1}^n \bar \phi_\alpha \left( E-H_0\pm i \epsilon
\right) \phi_\alpha  \pm i \frac{\lambda}{ 2} \left(
\sum_{\alpha=1}^n \bar \phi_\alpha \phi_\alpha \right)^2 \right].
\label{Sreplica}
\ee
Calculating the Green's function \rf{eq:effectrepl} now
involves doing the functional integral at arbitrary integer $n$
and then analytically continuing the answer
to $n \rightarrow 0$.
In fact, the same comments as those given after (\ref{GadvGret})
in the SUSY context apply here, and a duplication of
the so-far introduced variables is required
for the quantum particle in a random potential,
but we refrain here from writing out the details.
(The low energy effective theories are in fact
nonlinear sigma models\cite{KaneStone}, both in the replica
and the SUSY descriptions.)

The replica method is easily used in perturbative calculations,
where the number $n$ of copies typically appears in the form of a
polynomial in $n$, in any order in perturbation theory. This is
easily, and unambiguously, continued to $n\to 0$. In the context of
a  nonperturbative analysis, one would, at least naively, need a
critical theory for all  (large) integer values of $n$, each of which would
have a central charge $c(n)$. This may (but typically will not
uniquely) determine an analytic continuation into $n\to 0$. Such
an approach is known not to  be feasible for the 2D theory describing
the Integer Quantum Hall plateau transition discussed above. On
the other hand, the dilute self-avoiding polymer problem is
known to be described (in any dimensionality $d$),
due to P. G. deGennes\cite{DeGennes},
as the $n\to 0$ limit of the replica analog\,\footnote{\,Which bears
the same relationship to (\ref{Sreplica}),  that
(\ref{eq:actionsusyParisiSourlas}) has with
(\ref{eq:actionsusy}).} of the SUSY action
(\ref{eq:actionsusyParisiSourlas}).
 In $d=2$ dimensions, a number of
properties of this replica action can be obtained
exactly\cite{LoopModels,SaleurDuplantierLoops}
in the continuous range $-2 \leq  n \leq  +2$ of the parameter $n$,
and this model is often referred to as the $O(n)$ model.
A similar analysis and corresponding results exist also  for the 2D $q$-state Potts model
in the continuous parameter range $0 \leq q \leq 4$.
The $q \to 1$ limit of the $q$-state Potts
is known\cite{RevModPhysWuPotts}
to describe percolation.

Although the SUSY technique is better controlled than the replica approach,
it is limited to non-interacting random systems.  This is because the
SUSY technique is based crucially on the ability to represent the inverse partition
function, such as \rf{eq:part}, in terms of a fermionic
functional integral. This is only possible if the original problem
without disorder did not contain interactions (non-Gaussian terms).
A much-studied example
of a disordered classical  2D statistical mechanics system which {\sl is interacting}
is provided by the random-bond $q$-state Potts model.  It can
be analyzed\cite{qPotts} with the help of the replica trick in
an expansion in $(q-2)$ about the Ising case ($q=2$).

\vskip .1cm

This paper contains attempts by the authors to understand
in more detail conformal field theories at central charge $c=0$.
Our prime motivation arises from the desire to understand
better the structure of CFT underlying two-dimensional
disordered critical points. There is a significant number of
such critical points which are of great physical
interest but which are typically poorly understood.
(Some have been mentioned above.)
In particular, we give here a pedagogical
and detailed exposition of results
which appeared earlier in Ref.~\cite{Gurarie2002},
but we also present a variety of new, so-far unpublished results.

Specifically,  we review certain unusual features,
which distinguish $c=0$ conformal theories
from ordinary, say unitary CFTs.
One of the most dramatic
such features is the indecomposability, or
`logarithmic' structure which  typically (except in certain special
cases, including affine  current algebras) appears in the identity representation
of the Virasoro algebra at $c=0$.
This manifests itself through the appearance of a so-called
 `logarithmic partner' $t(z)$ of the stress energy tensor.
Moreover, $c=0$ CFTs possess a
novel `anomaly' number sometimes
denoted by $b$,  which plays, in some sense,
a role similar to the central charge in $c \not =0$ theories:
the parameter $b$ may be used to distinguish different $c=0$ theories.
These general properties of a $c=0$ CFT are discussed in
Section \ref{SectionConformalFieldTheoryatczero}, where we also
motivate and derive the fundamental OPE between the (ordinary) stress tensor
and its logarithmic partner.

An important role is often played in CFT by
so-called null-vectors (or: singular vectors).
These are Virasoro
descendants which are themselves primary.
They are known to occur
when primary operators have conformal weights contained in
the Ka\v{c} table (here at $c=0$). It is important
to know if such a null-vector can be set to zero,
because in that case correlation functions
involving the Ka\v{c}-table operator will satisfy
differential equations, which makes them easily
computable.  While in an ordinary (unitary) CFT null-vectors
are always set to zero, this is not
necessarily the case in a nonunitary theory,
like a $c=0$ CFT.

In Section \ref{LogarithmicAlgebra} we make a connection between
the `anomaly' number $b$ and Ka\v{c} null-vectors. Interestingly,
the  number $b$ controls the question as to whether certain
Ka\v{c}-table null-vectors vanish identically or not, and hence
whether certain correlation functions satisfy the corresponding
differential equations. We discuss
%, in turn,
the cases of Ka\v{c}-table operators with nonvanishing two-point
functions, first those with nonvanishing, and subsequently those
with vanishing conformal weights.

In Section \ref{SectionCriticalDisordered Systems} we
review aspects of   critical disordered
systems described  by the supersymmetry method.
In these theories,
a partner of the stress tensor, of the kind that appeared
in Section \ref{SectionConformalFieldTheoryatczero}
entirely from considerations of conformal symmetry,
emerges naturally  on grounds of supersymmetry.  Moreover, a pair of
conformal weight-two fermionic operators appears together with the
stress tensor and its partner in the same supersymmetry multiplet.

In Section \ref{SectionExtendedStressTensorMultiplet}
we show that based purely on conformal symmetry considerations,
there must exist additional fields,
besides the stress tensor $T(z)$ and its partner $t(z)$,
whose  holomorphic parts have conformal weight two,
if $t(z)$ is `logarithmic'.
Under the only assumption that these additional
fields form a fermionic (anticommuting) pair, we show that the
(holomorphic) OPEs between all these weight-two fields are automatically
covariant under a global $U(1|1)$ SUSY.

We end the main part of the paper by comments
and speculations about a possible extended
chiral symmetry in $c=0$ CFT, based on
the notions developed here.

Three appendices provide a number of technical details.
We show in Appendix A that the anomaly number $b$
must be unique in a given theory
(as discussed at the beginning of
Subsection \ref{SubSectionOperatorsWithNonVanishingDimension}).
Appendix B addresses details of the computation of the OPE of descendant operators
in the $c=0$ CFT
possessing the logarithmic features discussed in this
paper. We also demonstrate in Section \ref{SectionSubtractionLogs}
the  complete subtraction of logarithms to all orders
in the OPE \rf{single}.  Appendix C addresses certain details
referring to the footnote below \rf{littlelOneBigLOne}
in Subsection \ref{SubSectionOperatorsWithNonVanishingDimension}.

\section{Conformal Field Theory at \boldmath{$c=0$}}
\label{SectionConformalFieldTheoryatczero}

\subsection{$c \rightarrow 0$ catastrophe}
\label{Headingcatastrophe}
Conformal field theories (CFTs) with central charge $c=0$ are very different from those
with $c\not = 0$. Consider a CFT
with  central charge $c$ and a primary scalar operator  $A(z,{\bar z})$
with left/right conformal
weights $(h, {\bar h})$, $h={\bar h}$, and
nonvanishing\,\footnote{\,Operators with vanishing two-point function
appear naturally in nonunitary CFTs as members of
a logarithmic pair (see e.g. Eq.~(\ref{LogarithmicPair}) below),
and may perhaps be best discussed within this framework.\\[-3mm]~}
two-point function. We choose to consider operators whose two-point functions
are normalized to
unity,\footnote{\,We denote by $A^\dagger(z, {\bar z})$ the operator which is
conjugate (more generally: `dual')
to ${A}(z, {\bar z})$, i.e. the one with the property that the OPE
of $A$ with $A^\dagger$ contains the identity operator.  Our notation is
understood to include, of course, the special case where $A^\dagger=A$.\\[-3mm]~}
\be
\VEV{A(z,{\bar z}) A^\dagger(0,0)}= {1 \over z^{2 h}{\bar z}^{2
h} }\ .
\label{corrzbarz}
\ee

The operator product expansion (OPE) of this operator with its conjugate is
known\cite{ExplanationOPEBPZ} to be given by
\begin{eqnarray}
&&
A(z,{\bar z}) A^\dagger(0,0)
%\nonumber\\[3mm]
%&&
=\frac{1}{ z^{2 h}}
 \left( 1+ \frac{2 h}{  c} z^2 T(0) + \dots \right)
\frac{1}{ {\bar z}^{2  h}}
 \left( 1+ \frac{2 h}{  c} {\bar z}^2 {\bar T}(0) + \dots \right)
\nonumber\\[1mm]
&&\hskip 3cm  + \,
 {\rm other \ primaries}\,,
\label{AAOPEnonChiral}
\end{eqnarray}
where $T(z)$, and ${\bar T}(\bar z)$ are the holomorphic and antiholomorphic
components of the stress-energy tensor of the theory.  Here, `${other \ primaries}$'
denotes possible contributions to the OPE
from primary operators other than the identity operator.
From now on, in the rest of this paper,
we will focus entirely, as is customary,
on the holomorphic dependence, with the understanding
that a suitable `gluing' with the antiholomorphic
dependence has to be performed at the end to obtain bulk correlation functions.
With this understanding, we write the OPE
(\ref{AAOPEnonChiral}) as
\begin{equation}
\label{AAOPE}
A(z) A^\dagger(0) =\frac{1}{
z^{2 h}} \left( 1+ \frac{2 h}{  c} z^2 T(0) + \dots \right)
+ ... \ ,
\end{equation}
where $A(z)$ denotes in the usual way the `chiral (holomorphic) part' of the
operator $A(z, {\bar z})$.

This result, well known and general, cannot hold true in a CFT
with vanishing central charge. Indeed, a
direct limit $c \rightarrow 0$ in \rf{AAOPE} is not possible. We
call the $1/c$ divergence in \rf{AAOPE} a $c \rightarrow 0$
catastrophe.

To understand one way\,\footnote{\,There are two more ways
in which the $c=0$ catastrophe can be
resolved\cite{Gurarie1999,Cardy2001}:
(i) by operators with vanishing two-point functions
which may often naturally  be  thought of as  members of
a logarithmic pair \cite{Cardy2001},
or (ii) by operators with vanishing conformal weight (to be discussed
in Subsection (\ref{SubSectionOperatorsWithVanishingDimension})
below).}
how this catastrophe can get resolved\cite{Gurarie1999},
let us first consider the following example. Take a combination of two
 non-interacting
CFTs,
%conformal field theories,
one with central charge $b$, and one
with central charge $-b$.  We call their respective stress-energy tensors
$T_b(z)$ and $T_{-b}(z)$ which satisfy the well known OPEs
\be
\label{TbTb} T_b(z) T_b(0) =  {b/2\over z^4} + {2 T_b(0)\over z^2}
+ {{T'}_b(0)\over z} + ...\ , \qquad ({\rm and} \ \ b \to -b)
\ee
where prime denotes the derivative $\partial/\partial z$.
 The total stress-energy tensor is
$T(z)=T_b(z)+T_{-b}(z)$ and the total central charge $c=b+(-b)=0$.

A primary operator $A(z)$ of such a factorized theory would also be a product
of two operators, one in the theory with positive central charge,
and the other in the opposite theory.  The OPE of such an operator with
its conjugate can easily be found from
(\ref{AAOPE}),
$$
 A(z) A^\dagger(0) = \frac{ 1}{ z^{2 h}} \left(
1+\frac{h}{b}\, z^2 \left(T_b-T_{-b} \right) + ... \right) + ...\ .
$$
The problem of $c \rightarrow 0$ is now resolved, but the
resolution did not come for free. We now have to introduce a new
field \be t(z) \equiv T_{b}(z)-T_{-b}(z) \ee with conformal weight
$=2$, which is different from the stress-energy tensor
$T(z) = T_b(z) + T_{-b}(z)$ of the system.
This field will now always appear in such OPEs in the form
\be
\label{AAtOP}
 A(z) A^\dagger(0) =
\frac{1}{ z^{2 h}} \left( 1+\frac{ h}{ b}\, z^2 t(z) + ...
\right) + ...\ .
\ee
Continuing with our factorized theory, all OPEs between
the fields $T(z)$ and $t(z)$
 are easily computed from those of the factors
given in (\ref{TbTb}),
\begin{eqnarray}
\label{FactorizedTT}
T(z) T(0) &=&  {2 T(0)\over z^2} + {T'(0)\over z} + ... \ ,\\[2mm]
\label{nonlogTt}
T(z) t(0) &=& \frac{b}{z^4}
+ \frac{2 t(0)}{z^2} + \frac{t'(0)}{z} + ... \ , \\[2mm]
\label{eq:nonlogtt}
t(z) t(0) &=& \frac{2 T(0)}{z^2} + \frac{T'(0)}{
z} +  ... \ .
\end{eqnarray}
Note that the first equation reminds us
of the fact that at central charge
$c=0$ the stress tensor $T(z)$ is a primary field with vanishing
two-point function.

We would  now like to generalize this analysis to theories which
no longer  factorize into two non-interacting theories with equal
and  opposite central charges.  Based on our discussion to be
given below, we suggest that in any $c=0$ CFT a field of conformal
weight two, which we also denote again by $t(z)$, appears and that
it enters the OPEs of primary operators with nonvanishing
two-point function as in \rf{AAtOP}, thus resolving the
$c\rightarrow 0$ catastrophe. It follows from \rf{AAtOP} that $L_2
t=b$ (see e.g.  (\ref{LnWitht}) and (\ref{DetailsOPEAhAh}) of
Appendix B), which fixes the leading term in the OPE of $T(z)$
with $t(0)$ to be:
\begin{equation}
\label{TtOPENoLogs}
T(z) \,t(0) = {b \over z^4} + ...\ .
\end{equation}
However,  as far as the next order terms in this OPE
are concerned, they may or may not coincide with the
expansion given in  \rf{nonlogTt}.
 This is discussed in depth below in
(\ref{Ttgen}).

A relatively large class of  nontrivial theories realizing the (`nongeneralized')
OPEs
(\ref{AAtOP}), (\ref{nonlogTt}), (\ref{eq:nonlogtt})
are affine (or: Ka\v{c}--Moody) current algebras
with supergroup (or: `Lie superalgebra') symmetry, having
%with
central charge $c=0$.
One can show that
a pair of
{\it chiral} fields
$t(z, {\bar z}) = t(z)$
and
${\bar t}(z, {\bar z}) = {\bar t}({\bar z})$
with the properties discussed above, always
appears in these theories\cite{Gurarie1999}.  These
can be found as  expressions quadratic in
(Noether)
currents, and
transform under the supergroup symmetry as the
`top component' of an indecomposable multiplet of
stress-energy tensors (we will discuss such multiplets in more
detail in Section (\ref{SectionExtendedStressTensorMultiplet})
below).
The field $t(z)$ appears on the right-hand side of various OPEs
such as e.g. (\ref{AAtOP}),
and obeys \rf{nonlogTt}, (\ref{eq:nonlogtt}).
The number `$b$' becomes a property of
%parameter characterizing
the particular
affine (Ka\v{c}--Moody)
current algebra.

 However observe that, if $t(z)$ satisfies Eqs.~\rf{AAtOP}, \rf{nonlogTt}, and
\rf{eq:nonlogtt}, the algebra which $T$ and $t$ form
  becomes trivial, in the following sense.
Indeed, by reversing the arguments given above,  we can choose \be
\label{Diagonalization} T_b=(T+t)/2, \qquad T_{-b}=(T-t)/2 \ee
to re-diagonalize these equations and bring them into
 the form of two independent (commuting)  Virasoro algebras, with central charges
$b$ and $-b$, respectively.  From this point of view,
 affine (Ka\v{c}--Moody) Lie-superalgebras with $c=0$ are nothing but tensor
products of two non-interacting
CFTs
with equal and opposite central charges.

Quite remarkably, however, the OPEs \rf{AAtOP},  \rf{nonlogTt}
 and \rf{eq:nonlogtt} are but a special case of
a {\it more general set of  OPEs} at $c=0$,
 to be given in \rf{AA}, (\ref{Tt}) and \rf{txt} below.
We will now proceed to study theories with this more general form of OPE.

\subsection{Logarithmic partner $t(z)$ of the stress tensor $T(z)$ }
\label{SectionLogarithmicPartnerofStressTensor}

\begin{figure}[tbp]
 \centerline{\epsfxsize=3 in \epsfbox{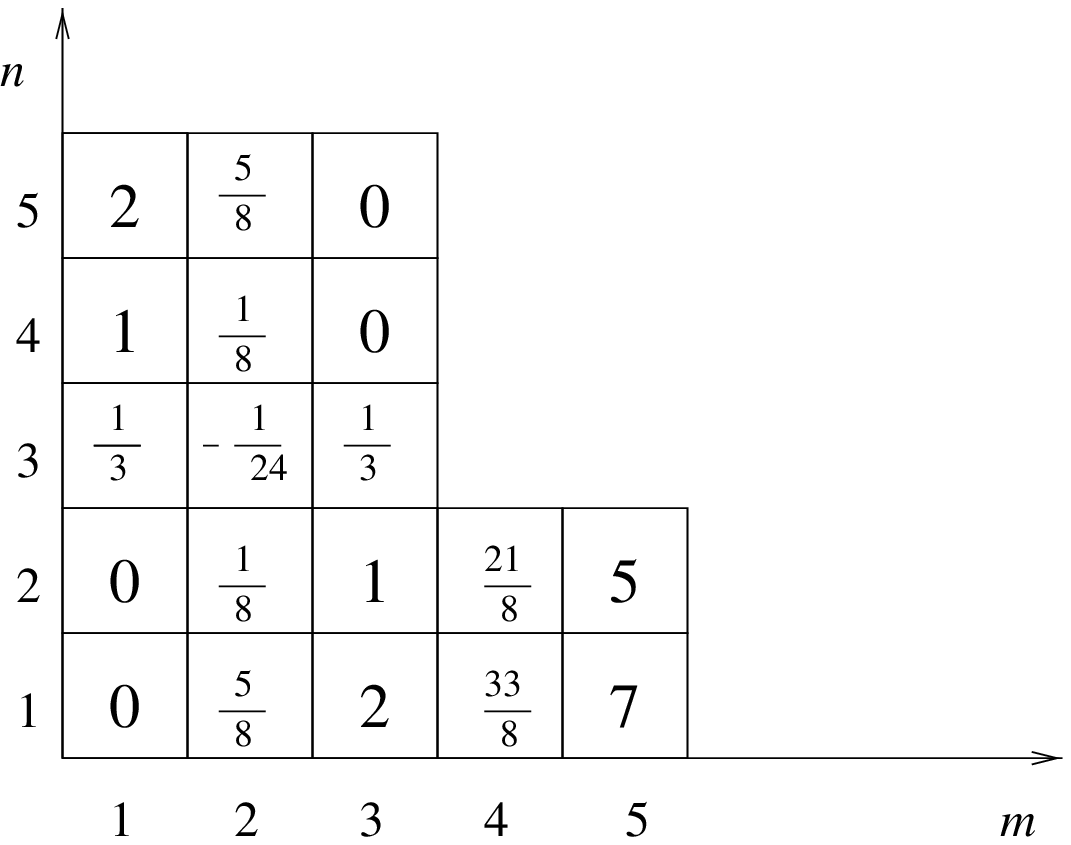}}
\caption {Some of the first few operators of the Ka\v{c} table at
$c=0$ } \label{fig:Kac}
\end{figure}

A special set of primary operators, the so-called Ka\v{c}-degenerate
operators, have conformal weights which lie on a two-dimensional
grid, usually referred to as the Ka\v{c} table. It is well known that
in conventional CFTs chiral (=holomorphic) correlation functions
involving at least one such `Ka\v{c}-degenerate' operator
satisfy\,\footnote{\,%
Even though this is
certainly the case in `conventional' CFTs (as opposed, e.g.,
to $c=0$ theories),
as discussed in  Ref.\cite{Belavin1984}, this issue is, as we
will see, more delicate for $c=0$ theories; see the discussion
following (\ref{Virasorofiveeights}) below.\\[-3mm]~} certain differential
equations \cite{Belavin1984}. Solving such differential equations
for the   (chiral) four-point functions (conformal blocks),
 provides a way to find the OPEs of primary operators. For further
reference we provide in
Fig.\,\ref{fig:Kac} a list of the first few operators of the
Ka\v{c} table at $c=0$.

Moreover, it is well known\cite{Belavin1984} that, due to global
conformal invariance, the (chiral)  four-point function of a primary
operator\,\footnote{\,For simplicity of presentation
we have chosen here  all four operators to be equal
and $A^\dagger=A$.}
can be expressed in terms of a single function $F(x)$,
\begin{equation}
\label{ChiralFourPtFct}
 \VEV{A(z_1) A(z_2) A(z_3) A(z_4) }
=  {1 \over (z_1-z_2)^{2 h} (z_3-z_4)^{2 h}}\, F(x)\, ,
\end{equation}
 where $ x$     denotes   a   cross-ratio
\begin{equation}
\label{DEFCrossRatio}
 x=
{(z_1-z_2) (z_3-z_4) \over (z_1-z_3) (z_2-z_4)}\,.
\end{equation}

Consider the ordinary differential equation
for the function $F(x)$, associated
with an operator $A$ belonging to the Ka\v{c} table.

In conventional CFTs (which have $c \not = 0$),
there is one solution of that equation which is of the form
\be
\label{IdentitiConfBlockNormal} F(x) = 1+  \alpha_0 \  x^2 +...
\ ,
\ee
(with some constant $\alpha_0$) corresponding to the
OPE \rf{AAOPE}. The function $F(x)$ with this expansion is usually
referred to as the  {\it identity conformal block} of the chiral four-point
function given in (\ref{ChiralFourPtFct}).

The situation at $c=0$ is far more complex, however. By
investigating the corresponding differential equation, it can be
directly verified that for all the operators from the first two
rows or from the first column of  the Ka\v{c} table in
Fig.~\ref{fig:Kac} (except for those with vanishing conformal weight,
discussed separately below),
 the small-$x$ behavior of the
identity conformal block is
\be
\label{expand}
F(x) = 1+ \alpha \
x^2 \log(x) + ...
\ee
in contrast to
(\ref{IdentitiConfBlockNormal}). It turns out that the other
operators of the Ka\v{c} table, which lie deeper in its interior (i.e.
beyond the first two rows or the first column), have even more
complicated identity conformal blocks \cite{Momo}. We will not
consider them in this paper, however.

The appearance of logarithms in a correlation function at a critical point,
as on the right-hand side of \rf{expand}, is characteristic of theories with so-called
 logarithmic operators \cite{Gurarie1993}. In this particular case,
the relevant logarithmic operator has conformal weight two, the
same weight as that of the stress tensor $T(z)$. Based on these
considerations we are led to suggest the following contribution to
the identity operator appearing in the OPE between any two primary
operators with nonvanishing two-point
function,\footnote{\,%
We do not suggest that (\ref{AA}) is necessarily an appropriate
OPE for Ka\v{c}-table operators.  Indeed, arguments
given in Ref.\cite{Cardy2001} would indicate that bulk Ka\v{c}-table
operators for percolation and for self-avoiding polymers
have vanishing two-point functions,
and this would not lead to the OPE (\ref{AA}).}
\be
\label{AA}
A(z)
A^\dagger(0)= {1 \over z^{2 h}} \left( 1+{h \over b} z^2 \left[
t(0) + \log(z) T(0) \right] + \dots \right)
 +   {\rm other \ primaries},
\ee where the ellipsis denotes higher descendants of the identity
operator, and `${other \ primaries}$' denotes contributions to
this OPE from operators other than the identity.
 This OPE generalizes Eq.~\rf{AAtOP}.
(For more details  about the structure of this OPE see
Sections \rf{AppBTheOPEofTwoPrimaryOperators} and
 \rf{SectionDetailsOPEAtwoAtwo} of Appendix B.)

In order to understand  why an OPE of this kind would give rise
to the logarithms in the conformal block (\ref{expand}),
 first recall\cite{Kogan1996,Gurarie1993}
that, in general, two (quasiprimary\cite{Belavin1984})
 operators
$C(z)$ and $D(z)$ of conformal weight $h$
are said to form a {\it logarithmic pair}, if
 the dilation operator $L_0$ does not act diagonally, but  in
`Jordan block' form, i.e.
\be
\label{LogPair}
L_0 C=h C\,,  \ \ \ \   L_0 D = h D + C\,.
%\qquad \ \ \ {\bar L}_0 C={\bar h} C,  \ \  {\bar L}_0 D = {\bar h} D + C
\ee
Global conformal invariance then enforces
the following form of the two-point
functions:
\begin{eqnarray}
\nonumber
\VEV{C(z)C(0)}&=&0 \,,\\[2mm]
\nonumber
\VEV{C(z)D(0)}&=& {a_1\over z^{2h}}\,,\\[1mm]
\label{LogarithmicPair}
\VEV{D(z)D(0)}&=&  {-2 a_1  \ln z + a_0 \over z^{2h}}\,.
\end{eqnarray}
Once the normalization of the operator $C$ is given, the
normalization of the operator $D$ has been fixed by requiring
the appearance
of the {\it same} coefficient $a_1$ in the second and in the third
equation.  The arbitrary constant $a_0$
arises from the freedom to redefine the operator $D$ by addition
of $C$ with an arbitrary coefficient, without changing the OPE's
above.

The (quasiprimary) operators $T(z)$ and $t(z)$ appearing in (\ref{AA}) form
precisely such a logarithmic pair, as we will see in (\ref{ttcor}) below.
 One now verifies immediately, upon identifying
$C \to T, D \to t, h \to 2, a_1 \to b$,
that the OPE (\ref{AA}) leads to the logarithm in  the small-$x$
 expansion (\ref{expand})
 of the identity conformal block of the chiral four-point function
upon using (\ref{LogarithmicPair})
(here we consider $A^\dagger= A$ for simplicity).
Furthermore, the coefficient $\alpha$ in (\ref{expand}) is
then fixed to be
\be
\label{hSquareOverb} \alpha = {h^2 \over b}\,.
\ee

The operator $t(z)$ we introduced in this manner in \rf{AA} fulfills the same
role as the operator $t(z)$ in (\ref{AAtOP}) of the previous
section. It makes sure the limit $c\rightarrow 0$ in \rf{AAOPE}
makes sense. However  at the same time, it is also
responsible for the logarithms in \rf{expand}, and therefore it is
a logarithmic operator. An OPE between the $c=0$ stress tensor
$T(z)$ and the new operator $t(z)$, which generalizes
(\ref{nonlogTt}), and which causes
$(T(z), t(z))$ to become a `logarithmic pair', is (as we will see shortly)
\be
\label{Ttgen}
T(z) t(0) =  {b \over z^4} + {2 t(0) +  \lambda T(0) \over z^2} +
{t'(0) \over z} +\ldots\ ,
\ee
where the parameter $\lambda$ is
arbitrary.  $\lambda=0$ corresponds to the previous, `nonlogarithmic OPE'
\rf{nonlogTt}, which implies (\ref{eq:nonlogtt}), and this can be
re-diagonalized as in the previous Section. On the other hand,
nonzero $\lambda$ corresponds to a new, `logarithmic OPE', as we
now explain. First, when $\lambda \not = 0 $, we can redefine
$t(z)$ and $b$ by dividing by $\lambda$ and
arrive at\,\footnote{\,%
The various terms are easy to understand.
The leading term is fixed by the considerations of \rf{TtOPENoLogs},
recalling that both OPEs, \rf{AAtOP} and \rf{AA}, imply $L_2 t =b$
(Appendix B, (\ref{LnWitht}) and (\ref{DetailsOPEAhAh})).
There is no $1/z^3$ term because $t$ is quasiprimary. The next order,
$1/z^2$-term is fixed by the `logarithmic' condition \rf{LogPair},
and by the conformal weight $h=2$ of $t(z)$.}
\be
\label{Tt}
 T(z) t(0) =  {b \over z^4} + {2 t(0) +  T(0) \over
z^2} + {t'(0) \over z} + ...\ .
\ee
This OPE generalizes the OPE \rf{nonlogTt}  derived for the
special (factorized)  situation considered in the previous Section.

The OPE \rf{Tt}  fixes the following set of (holomorphic)
two-point correlation functions (compare  with (\ref{LogarithmicPair}), recalling
that the normalization of the stress-tensor $T(z)$ is fixed),
\begin{eqnarray}
\nonumber
\VEV{T(z)T(0)}&=&0 \,,\\[2mm]
\nonumber
\VEV{T(z)t(0)}&=& {b\over z^{4}}\,,\\[2mm]
\label{ttcor}
\VEV{t(z)t(0)}&=&  {-2 b\ln z + \theta \over z^{4}}\ , \quad
(\theta~\mbox{is a \ constant})\,.
\end{eqnarray}
The first of these equations comes from the OPE
\rf{FactorizedTT} which remains unchanged,
independent of whether $t(z)$ satisfies the
previously discussed
`non-logarithmic' OPE \rfs{nonlogTt}, or its `logarithmic'
generalization \rf{Tt} which we are currently considering.
The second equation follows directly from the OPE \rf{Tt},
while the third
equation can be computed by imposing global conformal
invariance on $\VEV{t(z)t(0)}$ (see Appendix A for more details).
Hence, the `logarithmic OPE' in \rf{Tt}
is directly responsible for the logarithm appearing in the third of
(\ref{ttcor}), and thereby, for the logarithm in the conformal
block (\ref{expand}).

As mentioned above, the constant $\theta$ remains undetermined.
The occurrence of such undetermined constants is common
in the theory of logarithmic operators \cite{Gurarie1993,Kogan1996}, and
is related to the fact that $t(z)$ can be redefined as in
\begin{equation}
\label{redefine}
t(z) \rightarrow t(z) + \gamma T(z)
\end{equation}
with an arbitrary coefficient $\gamma$.
This redefinition does not affect any of the OPEs discussed in this
Section.

With a {\it logarithmic} operator $t(z)$ satisfying \rf{Tt} the theory is
no longer equivalent
to two commuting Virasoro algebras at $c \not =0$
(the last of (\ref{ttcor}) is an obstruction to the diagonalization
performed in (\ref{Diagonalization})). On the other hand, \rf{Tt} is the most
general OPE which a conformal weight-two (quasiprimary)
operator $t(z)$ with $L_2 t=b$ can satisfy
(recall the footnote preceeding Eq.~\rf{Tt}).
% following Eq.~\rf{Ttgen}
Therefore, we postulate that \rf{Tt} is
realized in all CFTs with central charge $c=0$,
except for those which simply factorize as in Subsection
(\ref{Headingcatastrophe}).

Finally, once the OPE \rf{Tt} between $t$ and $T$ is known, it is possible
to construct
(the contribution of the identity operator to)
 the OPE of $t$ with itself, which generalizes
\rf{eq:nonlogtt}. We can do
 this
by taking
the
most singular term of this OPE
from the correlation function $\VEV{t(z) t(0)}$ computed in
\rf{ttcor}.
This function contains an ambiguity related to the possibility to redefine
$t(z)$ according to \rf{redefine}. In what
follows, we fix this ambiguity by setting $\theta=0$ in
\rf{ttcor}. The most singular term entails all other terms of the
OPE, which becomes
\begin{eqnarray}
\label{txt}
t(z) t(0) &=& -{2 b \log(z) \over
z^4} + {t(0) \left[ 1-4 \log(z) \right] - T(0) \left[ \log(z)+2
{\log^2(z)} \right] \over z^2}
\nonumber\\[1mm]
 &&+ {t'(0) \left[ 1-4 \log(z)
\right] - T'(0) \left[ \log(z)+2 {\log^2(z)} \right] \over 2 z}+
...\, ,
\end{eqnarray}
where the ellipsis denotes
higher order terms, as well as contributions from primary operators
other than the identity operator.  The technique for
reconstructing
entire OPEs such as \rf{txt}
from their most singular terms
is well known and is described in Ref.~\cite{Belavin1984}, although its
application to the theory with logarithms has not often been
discussed in the literature.
Briefly, it consists of the following
steps (more details can be found in Appendix B, especially
Sections \ref{TheOPEgztzero} and \ref{DetailsofTheOPEtztzero}).
First we have to derive the commutation relations between the
Virasoro generators $L_n$ and $t$, from \rf{Tt}.  This yields
%\begin{eqnarray}
\begin{equation}
[L_n, t(z)] %&=&
=\Big( z^{n+1} \dd{z} + 2 (n+1) z^n
\Big) t(z)
%\nonumber\\[2mm]
%&+&
+(n+1) z^n T(z)+ {b\over 3!} (n^3-n) z^{n-2}\,.
\label{eq:com}
\end{equation}
%\end{eqnarray}
Then we
apply $L_n$ with $n\ge 0$ to \rf{txt}.
On one hand, $L_n$ can
be applied to the right-hand side of \rf{txt} directly. On the
other hand, we can use
$$[L_n, t(z) t(0)]=[L_n,t(z)] t(0) + t(z) [L_n, t(0)]$$
on the left-hand side,
substitute \rf{eq:com} into this expression, and use the
OPEs $t(z)t(0)$ and $T(z) t(0)$ to find relationships between various
terms in \rf{txt}. Ultimately, this allows
us to deduce, order by order in $z$,  all the
terms in \rf{txt} from its most singular term.

Equation \rf{txt} fixes the OPE $t(z)t(0)$ up to contributions of
other primary operators. $T(z)$ is a primary operator at $c=0$
and, since it already appears in Eq.~\rf{txt}, we can expect
on general grounds that  there could be a `stand alone' contribution of
the conformal block of the stress tensor to this OPE. This amounts
to
\be
  \frac{2  a \ T(0)}{z^2} +   \frac{ a \   T'(0)}{z} + ...
\ee
being added to the right-hand side of Eq.~\rf{txt} where $a$ is an
arbitrary coefficient. This was recently stressed
by I. Kogan and A. Nichols \cite{Kogan2003}. Additionally, one could imagine that
in specific $c=0$ CFTs which realize the logarithmic operator
$t(z)$, there could be contributions
from other primary operators on
the right-hand side of Eq.~\rf{txt}
(as already mentioned).  These, however, will play no further role in this paper,
until we arrive at
Subsection \rf{SectionCommentsOnAnExtendedAlgebra}.
% Section \ref{SectionExtendedStressTensorMultiplet}.

As we saw in the Introduction, in certain cases it is possible to think of
a  $c=0$ CFT as a limit of a continuous set of CFTs parametrized by
a parameter $n$, defined in an interval containing $n=0$,
where the central charge $c(n)\not = 0$ if $n \not = 0$, and $c(0)=0$.
In that case, we could ask how a partner $t(z)$ of the stress tensor
can appear in the limit $n
\rightarrow 0$, while it is definitely not present in the theory
at $n \not = 0$. The answer to that question was given
by J. L. Cardy in Ref.\cite{Cardy1999,Cardy2001}.

 We summarize by saying that the OPEs (\ref{Tt}) and (\ref{txt}), together with
the OPE of the stress tensor $T(z)$ with itself (which is unmodified,
and as in (\ref{FactorizedTT})) constitute the fundamental
equations of a CFT at central charge $c=0$, which does not factorize as in Subsection
(\ref{Headingcatastrophe}).

\section{Implications of the logarithmic \boldmath{$t(z)$}
and corresponding  \boldmath{$b$} on  \boldmath{$c=0$} Ka\v{c}-table operators
with nonvanishing two-point functions}
\label{LogarithmicAlgebra}

In this section we study the implications of the `anomaly' number
$b$ for null-vectors, associated with primary operators which have
conformal weights listed in the $c=0$ Ka\v{c} table,
 and nonvanishing two-point functions.\footnote{\,As already mentioned,
operators with vanishing two-point function may often naturally be
viewed as members of a logarithmic pair (see e.g. Eq.~%
\rf{LogarithmicPair} above).} For ordinary (e.g. unitary) CFTs
there is no issue, because the null-vectors are known to vanish
when inserted into any (chiral) correlation function with other
operators\cite{Belavin1984}. This step is no longer guaranteed to
be valid for the nonunitary theories discussed here.  In the first
part,  Subsection
\ref{SubSectionOperatorsWithNonVanishingDimension}, we demonstrate
how the `anomaly' number $b$ controls this issue, for primary
operators with certain nonvanishing Ka\v{c}-table weights (and
non-vanishing two-point functions). In the second part, Subsection
\ref{SubSectionOperatorsWithVanishingDimension}, we discuss
similar statements for Ka\v{c}-table operators with vanishing
conformal weight (and nonvanishing two-point functions).  A
convenient tool used in both subsections to address these
questions in a purely algebraic way, is a (partial) extension of
the Virasoro algebra by suitably defined Laurent modes of the
logarithmic partner $t(z)$ of the stress tensor.

\subsection{Operators with nonvanishing dimensions}
\label{SubSectionOperatorsWithNonVanishingDimension}

In view of the relation (\ref{hSquareOverb})
$$
 \alpha = {h^2 \over b}
$$
it may appear, at first sight, that a separate, and possibly
different value of the parameter $b$ could be associated with
different $c=0$ primary operators $A(z)$.
This would mean, that a given theory would
contain two (or more) different values of $b$, say $b \not = b'$.
But this would imply that there would exist two (or more) different
operators $t(z)$, say $t_b(z)$ and $t_{b'}(z)$,
each obeying the OPE \rf{Tt} with the
coefficient of the corresponding $1/z^4$ term equal
to $b$ and $b'$, respectively.
%But this would imply that there would exist
%several operators $t(z)$, each obeying the OPE \rf{Tt} with
%the corresponding value of $b$.
%Specifically, let us denote by  $t_b(z)$ and $t_{b'}(z)$ two
%different operators each obeying the OPE \rf{Tt} with the $1/z^4$
Then it is not difficult to see that the correlator
$\VEV{t_b(z) t_{b'}(0)}$ violates global conformal
invariance. (Details are given in Appendix A.)
 This means that
different values of $b$ cannot coexist in the same theory,
 and that  $b$ is a characteristic of any $c=0$ CFT.
Therefore, the question arises what value the number $b$
takes  in a given theory, and what conditions $b$ imposes on the properties
of the  $c=0$ CFT.
This is the question we address in  this Section.

We begin by considering a $c=0$ theory containing in its operator
content one or more primary operators whose conformal weight
appears in the Ka\v{c} table. If we were to assume that the
corresponding null-vector, implied by the Ka\v{c}-table conformal
weight,  can be set itself to zero,\footnote{\,See
(\ref{Virasorofiveeights}) below for an example, and a more
in-depth discussion of this issue.} then any  (chiral) four-point
function (conformal block) involving this operator would satisfy a
differential equation. For any primary operator with Ka\v{c}-table
conformal weight (and nonvanishing two-point function)  it would
hence be  possible to extract the coefficient $\alpha$ appearing
in \rf{expand} from the solution of the corresponding differential
equations for $F(x)$ (see (\ref{ChiralFourPtFct})). Therefore, in
view of (\ref{hSquareOverb}), this associates a value of $b$ with
any such Ka\v{c}-table operator.

In the following, we will obtain a purely algebraic way of
associating a number $b$ with operators in the $c=0$ Ka\v{c} table
which have nonvanishing two-point function, without explicitly
referring to the corresponding differential equation, or its
solutions.
% This method, then, permits us to discuss primary operators with Ka\v{c}-table conformal weights,
%irrespective of whether their two-point functions vanish or not.
Interestingly,  we find that different values of $b$ appear
in the $c=0$ Ka\v{c} table.
 For example, operators in the first two rows of the Ka\v{c} table have
$b=+5/6$, whereas operators in the first column have $b=-5/8$ (see
(\ref{deg}) below).

Since the value of the number $b$ is
unique in a given theory (as per our discussion
at the beginning of this section),
this implies
that only those Ka\v{c}-table operators
which have a given fixed value of $b$ can
give rise to differential equations
in a given theory.

In order to arrive at our algebraic determination of $b$ for
operators with conformal weights contained in the  $c=0$
Ka\v{c} table, we
will  first
%need to
establish the OPE between $t(z)$, and an arbitrary primary
operator $A_h(z)$ of conformal weight $h \not =0$ and nonvanishing
two-point function.

We start by determining the three-point correlation functions
involving these operators. It is well known that the three-point
correlation functions are completely determined by global
conformal invariance. By imposing global conformal invariance on
$\VEV{t(z) A_h(w_1) A_h(w_2)}$  one readily finds
\be \label{tAA}
\VEV{t(z) A_h(w_1) A_h(w_2)} = {h \log \left( {w_1-w_2 \over
(z-w_1) (z-w_2)} \right)  + \Delta \over (z-w_1)^2 (z-w_2)^2
(w_1-w_2)^{2 h - 2} } \,.
\ee
The coefficient $\Delta$ is
arbitrary and is not fixed by conformal invariance. In what
follows we set $\Delta =0$, which amounts to redefining $t(z)$ as
in \rf{redefine} in a suitable way. (Notice that this would not be
possible for operators $A_h$ with vanishing conformal weight,
$h=0$. Therefore we consider for now only operators $A_h$ with
nonvanishing conformal weight.)  Now consider expanding the
three-point function \rf{tAA}  for small $(z_1-w_1)$. One
immediately sees that the term
%proportional to
multiplying $\log (z_1-w_1)$ is
%equals
precisely  equal to
$<T(z) A_h(w_1) A_h(w_2)>$.
Moreover, an additional power series in $(z-w_1)$ appears
which  does not
%is not proportional to
multiply $\log (z_1-w_1)$.
All this is consistent with the following OPE,
\be
\label{tA}
t(z) A_h(0) = - T(z) A_h(0) \log(z) +
\sum_{n=0}^{+\infty}
\ell_{-n} A_h(0) z^{n-2}
\ \ + \ \Re(z)\,,
\ee
where potential noninteger powers of  the variable $z$ (but no logarithms)
are collected in a `remainder' denoted by $\Re(z)$.
This can be viewed as a definition of the operators
$\ell_{-n} A_h(0)$.
Furthermore, if the operator $A_h$ is
replaced by a more general (e.g.  nonprimary) operator,
negative powers of the index $n$ may appear  in the OPE
\rf{tA}.

Alternatively, the action of the operators $\ell_n$ on the operator
 $A_h(0)$ can be computed  as usual by contour integration from \rf{tA},
\be
\label{Defelln}
\ell_n =
{\hat {\bf P}}_h
\
\left [
  \
\oint  {dz \over 2 \pi i}  \ \biggl ( t(z) + \log(z) T(z) \biggr) z^{n+1}
\
\right ]
 \
{\hat {\bf P}}_h,
\qquad  (n \in {\bf Z}),
\ee
where ${\hat {\bf P}}_h$ is the projection operator on
all states of the Hilbert space of the CFT whose
conformal weights differ from the weight $h$ of the
operator $A_h(0)$ by an integer
(this projection operator
commutes with the operator $T(z)$; see e.g. \rf{ExpansionT}).

Formula \rf{Defelln}, together with the
definition of the conventional Virasoro generators
\be
L_n=\oint {dz \over 2 \pi i}   ~T(z) z^{n+1},
\ee
allows us to find the commutation relation $[\ell_n, L_m]$
in the familiar manner from the OPE \rf{Tt}. The result is
\be
\label{CommuatorsLnln}
[\ell_n, L_m] =
+(n-m)\ell_{n+m}
-m  L_{n+m}
+{b \over 6}\, n(n^2-1)\delta_{n+m, 0}\ .
\ee
It constitutes a (partial) generalization of the
commutation relations of the Virasoro algebra, which are given
at $c=0$ by
\be
\label{VirasoroAlgebra}
\left [
L_n , L_m
\right ]
= (n-m) L_{n+m}\,.
\ee
We now claim that
\be
\label{lnA}
 \ell_n A_h(0)=0\,, \qquad \qquad
{\rm for  \ all} \ \  n \geq 0 \ee for primary operators $A_h$
with nonvanishing two-point function.  First, it clearly follows
from the OPE \rf{tA} that this is true for  $n>0$. Moreover, we
can choose a definition of the operator $t(z)$ so that  $\ell_0
A_h(0)=0$; indeed, since $t(z)$ is defined up to addition of
$T(z)$ as in (\ref{redefine}), $\ell_n$ is also defined up to
addition of $L_n$, as in \rf{redefine}. By adding $L_0$ to
$\ell_0$ with a suitable coefficient, we can always\,\footnote{\,A
different such  `subtraction' will typically have to be performed
for each operator $A_h(z)$ separately. This will not affect the
arguments given below.} make $\ell_0 A_h(0)$ vanish as long as
$h=L_0 A_h(0) \not = 0$. From now on we assume that we have chosen
$t(z)$ in this way, while we consider the operator $A_h(z)$.

Using the newly derived commutators (\ref{CommuatorsLnln}),
the ordinary Virasoro algebra \rf{VirasoroAlgebra},
as well as (\ref{lnA}),
one finds that
\be
L_1 \left( \ell_{-1}- {1 \over 2} L_{-1}\right) A_h(0) = 0.
\ee
Therefore, $\left.(\ell_{-1}-{1 \over 2} L_{-1})|A_h\right\rangle$
can be called a {\sl null-vector}, following
Ref.~\cite{Belavin1984}.  Indeed, from now on we will set  this null-vector to zero, i.e.
\be
\label{l-L}
\ell_{-1} A_h(0) = {1 \over 2} L_{-1} A_h(0),
\ee
which is easily seen to be consistent with the correlation function \rf{tAA}.
Moreover, this is also consistent with
the general constraints of conformal symmetry imposed
on the OPE \rf{tA}  (see e.g. Section
\ref{SectionDetailsOPEtwithA} of Appendix B).

Any primary operator  whose conformal weight appears
in the Ka\v{c} table
has descendants which are themselves null-vectors. Specifically,
this means that there exist states $|\xi\rangle$,  constructed by
applying Virasoro lowering operators $L_{-m}$ \  ($m >0$),  to the
primary state $\left. A_h|0\right\rangle$, so that
$L_n|\xi\rangle=0$, for all $n>0$. However, now we have a (partial)
extension of the Virasoro algebra in hand, generated by $L_n$ and
$\ell_n$. Hence it is
%reasonable
natural to  ask if the null-vectors are annihilated by $\ell_n$ as
well as by $L_n$. Consider for example the primary operator with
conformal weight $h_{(2,1)}={5 \over 8}$ which is contained in the
$c=0$ Ka\v{c} table. We assume it has nonvanishing two-point
function.  Its  (Virasoro) null-vector is \be
\label{Virasorofiveeights} \left(L_{-2}-{2 \over 3}
L_{-1}L_{-1}\right) A_{5 \over 8}(0) \ee which means that this
expression is annihilated by $L_n$ with $n>0$. We are interested
in knowing if we are allowed to set the operator appearing in
(\ref{Virasorofiveeights}) {\it itself to zero}, when it occurs in
{\it any}  correlation function with other operators.  This is
important to know,  because  if true it would give
rise\cite{Belavin1984} to a differential equation  satisfied by
any conformal block involving the Ka\v{c}-table operator
$A_{5/8}$. In view of the nonunitarity of the present theory, it
is not obvious that (\ref{Virasorofiveeights}) itself can be set
to zero. (See the paragraph below,  containing
\rf{StressTensorPrimaryatCisZero}, for a related example.) Now we
observe that by applying the operator $\ell_2$ to
(\ref{Virasorofiveeights})  and by  using (\ref{CommuatorsLnln})
one finds \be \label{elltwoAppliedToNullVector}
%\ell_{+2} \bigl ( \Lambda A_{5/8}\bigr )(0)=
\ell_{+2}
 \left(L_{-2}-{2 \over 3} L_{-1}L_{-1}\right)A_{5/8}(0)=
\bigl ( b - 5/6 \bigr ) A_{5/8}(0) \ee which vanishes only if
$b={5 \over 6}$.   Furthermore, when  applying the
operator $\ell_1$ to (\ref{Virasorofiveeights}) we arrive, irrespective of
the value of ``$b$'', at
\be
\label{littlelOneBigLOne}
\left( \ell_{-1}-{1
\over 2}L_{-1} \right) A_{5 \over 8}(0),
\ee
which, as we have
already established in (\ref{l-L}), vanishes for all primary operators $A(z)$
with nonvanishing conformal weight. (Applying $\ell_{n}$ with
$n\geq 3$ is easily seen to always annihilate (\ref{Virasorofiveeights}).)

Hence we arrive at the important conclusion that {\it if} the
descendant (\ref{Virasorofiveeights})
 of a Ka\v{c}-table operator may be set to zero
in any correlation function with other operators
(which, we emphasize again,  implies\cite{Belavin1984}
the validity of the corresponding
differential equation involving this operator),
then $\ell_2$ and $\ell_1$ applied to
it must also
vanish,\footnote{\,Consider e.g. a correlation function
involving $t(z)$, which is known to appear
in OPEs between primary operators,
such as (\ref{AA}); see Appendix C for  further
elaboration.}
and this does not happen unless
$b={5\over 6}$.
This is a necessary condition determining the  value of $b$
 by simple algebra,  without solving
the  differential equations ensuing from the null-vector
condition.  On the other hand, if we took the route described at
the beginning of this subsection,  i.e. if we  solved the
differential equation for the four-point function of the operator
$A_{5 \over 8}(z)$
 (assuming it had a nonvanishing two-point function),
found the coefficient $\alpha$ defined in \rf{expand}, and
determined $b$ via (\ref{hSquareOverb}), it would also be ${5
\over 6}$. Therefore, the above steps establish a purely algebraic
way to determine for each Ka\v{c}-table operator
 with nonvanishing two-point function
 a value of $b$, for which
the null-vector (such as e.g.  (\ref{Virasorofiveeights})) may be
set to zero.  This, in turn, gives rise to the ensuing differential equation.

The notion of a nonvanishing null-vector
%at $c=0$
may be unfamiliar.  To illustrate  it,
%this,
let us give a related, but simpler example:
 at $c=0$ the stress tensor,  which is a descendant
of the identity operator (e.g. \rf{ExpansionT}),
is itself primary,
\be
\label{StressTensorPrimaryatCisZero}
L_{+2} L_{-2} {\bf 1}(0) =0\,,
\qquad
L_{+1} L_{-2} {\bf 1}(0) =0\,.
\ee
The stress tensor hence represents a null-vector
of the identity operator at level two.
But clearly, the stress tensor does not vanish,
even though its two-point function does.
And indeed, in analogy with \rf{elltwoAppliedToNullVector},
we obtain from \rf{CommuatorsLnln}
\be
\label{lPlusNStressTensorPrimaryatCisZero}
\ell_{+2} L_{-2} {\bf 1}(0) =
b  \ {\bf 1}(0)
\ee
which does not vanish (unless $b=0$ in this case).

While we lack a general result for $b$ based on the above method for
{\sl all}
 operators in the $c=0$ Ka\v{c} table, we have
 repeated this procedure for many of the operators of the
Ka\v{c} table and found the following pattern\cite{MonwheaJeng}:
\begin{eqnarray}
\nonumber
{\rm for} \
A_{(k,1)} \ \& \ A_{(k,2)}, \ k>1 \quad  {\rm (first \ two \  rows)}: \quad \ \ && \ \ b=+{5 \over 6}  \,,\\[1mm]
\label{deg}
{\rm for} \ A_{(1,k)}, \ k>2  \ \ \quad  {\rm (first \  column)} : \quad \ \ && \ \  b=-{5 \over 8}\, .
\end{eqnarray}
Here $A_{(m,n)}$ denotes the operator located in position $(m,n)$
of the Ka\v{c} table of Fig.~\ref{fig:Kac}.

In view of the uniqueness of the number $b$ in any given theory,
the appearance of  {\it different} values for $b$ in \rf{deg} has
important consequences. It means that, in a given $c=0$ theory,
only certain subsets of primary operators with conformal weights
given by the Ka\v{c}-table have null-vectors which vanish
identically (implying that the  corresponding conformal blocks
satisfy differential equations). To be entirely clear, but at the
risk of being repetitive, let us spell this out once more in
detail (all Kac-table operators mentioned below are assumed to
have nonvanishing two-point function).  Take for example the
operator $A_{(2,1)}$ of conformal weight ${5 \over 8}$, which we
previously denoted as $A_{5 \over 8}$. Conformal blocks involving
this operator satisfy the (second order) differential equation
associated with the null-vector of $A_{(2,1)}$ at the second level
{\it only if} $b={5 \over 6}$. On the other hand, take the
operator $A_{(1,3)}$ with conformal weight ${1 \over 3}$.
Conformal blocks involving this operator satisfy the (third order)
differential equation associated with the null-vector of
$A_{(1,3)}$ on the third level {\it  only if} $b=-{5 \over 8}$.
Therefore,  these two operators cannot give rise to the
corresponding differential equations simultaneously in the same
theory.  That does not mean that primary operators with `wrong'
conformal weights are necessarily forbidden in the same theory.
But it means that for `wrong' operators the null-vectors cannot be
set to zero, which implies that their correlation  functions would
not satisfy the differential equations which would otherwise
follow from these null-vectors (such as
(\ref{Virasorofiveeights})) according to the rules of
Ref.~\cite{Belavin1984}. For example, if $b={5\over 6}$, only the
(second order) differential equation associated with the
null-vector of $A_{(2,1)}$ at the second level can be valid.  But
the (third order) differential equation associated with the
null-vector of $A_{(1,3)}$ on the third level would not be valid.
In other words, a conformal block involving $A_{(1,3)}$ would not
satisfy this differential equation.
 Conversely, if $b=-{5 \over 8}$, only the
(third order) differential equation associated with the
null-vector of $A_{(1,3)}$ on the third level can be valid. But
the (second order) differential equation associated with the
null-vector of $A_{(2,1)}$ at the second level would not be valid.
In other words, a conformal block involving $A_{(2,1)}$ would not
satisfy this differential equation. Finally, if $b$ is not equal
to either of these numbers, neither of these  differential
equations will be satisfied by the conformal blocks involving
these operators.
(An explicit example illustrating these issues for the
Ka\v{c}-table operators $A_{3,1}$ and $A_{1,5}$ of
conformal weight two can be found in \rf{GclosedFormSolution}
below.)

This concludes our discussion of the operators with nonvanishing
conformal weights.

%\vskip 3cm

\subsection{Operators with vanishing dimension}
\label{SubSectionOperatorsWithVanishingDimension}

 Let us concentrate now on the operator $A_{(1,2)}$,
appearing in position $(1,2)$ of the
Ka\v{c} table in Fig.\,\ref{fig:Kac}.
A remarkable feature of this operator is that its conformal  weight
vanishes at central charge $c=0$,
even though the operator itself is different from
 the identity operator.
It is well known that this can happen in nonunitary
theories, and the operator $A_{(1,2)}$ plays a prominent role in
the theory of percolation \cite{Cardy1991}.

Consider first the correlation function on the left-hand side of
 \rf{tAA} where $A(z)$ is
now the zero-dimensional operator with $h=0$. In what follows, we
denote this operator by  $O(z)$, and we assume that it has
nonvanishing two-point function.  The right-hand side of \rf{tAA}
has been obtained from  global conformal invariance alone, and is
therefore certainly also valid when the conformal weight $h=0$. In
this case it reduces to \be \label{tOO} \VEV{t(z) O(w_1) O(w_2)}=
{\Delta (w_1-w_2)^2 \over (z-w_1)^2 (z-z_2)^2  }\,. \ee A
difference between \rf{tOO} and \rf{tAA} for the operator with
nonvanishing conformal weight is the absence of the logarithms.
Also, it is no longer possible to set $\Delta$ to zero by
employing \rf{redefine}.

Now consider how the OPE \rf{AA} changes when $h=0$.
It becomes
\be
\label{OO}
O(z) O(0) = 1 + C z^2 \bigl ( T(z) + ... \bigr) + ...
\ee
(as one might have expected from \rf{AAOPE}).
Notice the logarithms no longer appear, and a
%the
contribution of the stress-energy tensor conformal block
%shows up,
appears,  with an arbitrary coefficient $C$. The associativity of the
correlation function $\VEV{O(z_1) O(z_2) A_h(z_3) A_h(z_4)}$,
where $A_h$ is an arbitrary primary operator with conformal
weight $h$, requires $C={\Delta \over b}$.

Finally, consider how the OPE \rf{tA} changes.
It becomes
\be
\label{tO} t(z) O(0) = - (1-\epsilon) T(z) O(0) \log(z) + \sum_n
\ell_{n} O(0) z^{-n-2}
+...\,,
\ee
where $\epsilon$ is a new constant
which cannot be determined
from conformal invariance alone.

It follows from \rf{tOO} that $\ell_0 O(0)=\Delta O(0)$,
in contrast to the case of
operators with nonvanishing conformal weight $A(z)$,
for which we can arrange for $\ell_0 A(0) =0$ by a redefinition as
in (\ref{redefine}).
 From \rf{tO} something even more drastic
follows: the commutation relations of $\ell_n$ and $L_m$, when
they act on the zero-dimensional operator $O$, change from \rf{CommuatorsLnln}
to\,\footnote{\,%
A situation, where  the commutation relations depend on the operators on which
they act is encountered in other CFTs.  An  example of this is the
parafermion CFT\cite{Zamolodchikov1986}.}
\be
\label{l-L0}
[L_n,\ell_m] = {b \over 6}
(n^3-n)\delta_{n+m,0} + (n-m) \ell_{n+m} + \left( n+ \epsilon
\right) L_{n+m}\,.
\ee

Let us  try to use the commutators (\ref{l-L0})
to determine if the
vanishing of the
null-vector of $A_{(1,2)}$  at the second level imposes
any constraints on the parameters $b$, $\Delta$ and $\epsilon$.
In this case the null-vector is
$\left[ L_{-2}-{3 \over 2} L_{-1} L_{-1} \right]A_{(1,2)}$.
By applying $\ell_2$ and $\ell_1$ (as before), we
find that the following conditions need to be
satisfied, if the null-vector itself
can be set to zero,
\begin{eqnarray}
b &=& 5 \Delta\, ,
\nonumber\\[1mm]
 \Delta &=& {-5+7 \epsilon \over 12} \ .
\end{eqnarray}
Assuming various values of $\epsilon$ yields
the following values of $b$:
\begin{eqnarray}
\epsilon &=& 0 \ \rightarrow \ b=-{25 \over 12}\,,
\nonumber\\[1mm]
\epsilon &=&{1 \over 2} \ \rightarrow \ b=-{5 \over 8}\,, \nonumber\\[1mm]
\epsilon &=& 1 \ \rightarrow \ b={5 \over 6}\,.
\end{eqnarray}

Quite independently, the differential equation
 associated which the operator $A_{(1,2)}$
allows us to determine $\Delta$. Indeed, let
us calculate the (chiral) four-point correlation function
$\VEV{A_{(1,2)}(z_1) A_{(1,2)}(z_2) A_h(z_3) A_h(z_4)}$ where
$A_h$ is an arbitrary primary operator with conformal weight $h$.
Writing the second order differential equation which follows from
setting to zero the null-vector of $A_{(1,2)}$ at the second level,
we find  an identity  conformal block of the following form,
 \be
\label{AzeroAzeroAhAh}
\hspace{-0.1cm} ~\VEV{\!A_{(1,2)}(z_1) A_{(1,2)}(z_2)
A_h(z_3) A_h(z_4)\!}={1 \over (z_3-z_4)^{2h}} \Big( 1+{h \over 5}
x^2+ ... \Big)\, ,
\ee
 where, as before,
$$x=
{(z_1-z_2) (z_3-z_4) \over (z_1-z_3) (z_2-z_4)}\ . $$

Matching the coefficient ${1 \over 5}$ with the OPEs \rf{AA} and
\rf{OO} we find $C={1 \over 5}$ and consequently $\Delta={b \over
5}$. This is consistent with the purely algebraic results obtained
above. The value of $b$ cannot be determined in this way, however.
 This ends our discussion of operators with vanishing conformal
weight.

\section{Critical Disordered Systems}
\label{SectionCriticalDisordered Systems}

\begin{figure}[tbp]
\centerline{\epsfxsize=2 in \epsfbox{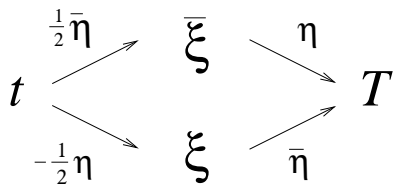}}
\caption {The multiplet of fields of conformal weight-two for the
supergroup U$(1|1)$.  $\eta$ and $\eta^\dagger$ are two fermionic
generators of this group. (In the figure, $\eta^\dagger$  is denoted by $\bar \eta$,
and $\xi^\dagger$  is denoted by $\bar \xi$.)}
\label{fig:mult}
\end{figure}

Consider a generic disordered system where the disorder average
can be performed using the  supersymmetry (SUSY) method,
 resulting in an action such as that given in
\rf{eq:actionsusy}.  A theory of this kind is always invariant with
respect to isotopic supersymmetry (`supergroup') transformations. For example, the
 action given in \rf{eq:actionsusy} is invariant under superunitary rotations
\begin{equation}
 \begin{pmatrix} \phi' \\ \psi'  \end{pmatrix} = U
\begin{pmatrix} \phi \\ \psi \end{pmatrix},
\end{equation}
where $U$ is a superunitary matrix. According to
Ref.~\cite{Gurarie1999} the stress tensor of such systems is
always a member of a  certain indecomposable
 SUSY multiplet. The number of fields in this
multiplet depends on the symmetry group of the system. Any such
theory must, however,  at least be invariant under a minimal
U$(1|1)$ SUSY, giving rise to a 4-dimensional (indecomposable)
multiplet of stress tensors denoted by
$T(z)$, $t(z, {\bar z})$, $\xi(z, {\bar z})$, $ \xi^\dagger(z, {\bar z})$
in Ref.~\cite{Gurarie1999}.
All four fields must have conformal weights
$(h, {\bar h}) =$  $(2, 0)$.
This multiplet, together with the action of the U$(1|1)$ generators
denoted (as in\cite{Gurarie1999}) by $\eta, \eta^\dagger, j, J$,
satisfying the relations
\be
\left [ j, \eta \right ] =  -2 \eta,
 \quad
\left [ j,  \eta^\dagger \right ] = +2 \eta^\dagger,
\quad
\{ \eta, \eta^\dagger \} =J,
\ee
 is depicted
in Fig.\,\ref{fig:mult}.
(The operators $(t, \xi^\dagger, \xi, T)$
transform in the same way as $(j, \eta^\dagger, \eta, J)$.)
If the SUSY is larger than this minimal $U(1|1)$, then this multiplet
will in general contain more fields, but the above four will always be
contained therein.

A special role is played by the `top' field $t(z,\bar z)$ of the
multiplet, displayed in Fig.  \ref{fig:mult},
whose OPE with the stress tensor $T$ was argued in Ref.~\cite{Gurarie1999}
 to satisfy (cf. Eq.~(\ref{TtOPENoLogs}))
\be
 \label{corrTt}
\langle T(z) t(w, \bar w)\rangle = {b \over (z-w)^4}\ ,
\ee
where the parameter $b$
counted\,\footnote{\,%
In a replica theory, the number $b$
basically corresponds\cite{Gurarie2002,Cardy2001,Kogan2002,Kogan2003}
to the  so-called `effective central charge' ${\partial/\partial n}_{|n=0} \  c(n)$,
where $c(n)$ is the central charge of $n$ coupled replicas, if only one uses
a slightly different normalization of the operator $t(z)$ such as in \rf{Ttgen}.}
the number of effective degrees of freedom of the disordered system.
Different disordered systems, all having central charge $c=0$, can be
distinguished by different values of $b$.  It was further suggested
in Ref.~\cite{Gurarie1999} that the field $t(z, \bar z)$ together with the
stress-tensor $T(z)$ should generate a certain  extension of the
Virasoro algebra, via their OPE. However, the most general form of
this OPE was not established.

At a critical point  a disordered system will typically have, as
mentioned, vanishing central charge $c=0$. According to the analysis
in Section \ref{Headingcatastrophe}, this implies the existence of an
operator with holomorphic part $t(z)$, designed
to avoid the `$c \rightarrow 0$ catastrophe'.
At this stage, upon comparing \rf{corrTt} with  \rf{TtOPENoLogs},
we are forced to identify the holomorphic part of
$t(z,\bar z)$ obtained in Ref.~\cite{Gurarie1999} by supersymmetry methods,
with $t(z)$ considered in Section \ref{Headingcatastrophe} of this paper.
Thus, at supersymmetric disordered critical points, a  partner of the
stress tensor is known to appear simply on grounds of (super-) symmetry.

An important  remark has to be made at this point.
We know from the
analysis of Ref.~\cite{Gurarie1999}
that in general, the operator $t(z, {\bar z})$ does not have to be holomorphic,
as indicated by the notation.
Although there exist theories such as Ka\v{c}--Moody super current algebras,
where it is certainly  holomorphic (but not logarithmic, as discussed
in Section \ref{Headingcatastrophe}),
$t(z, {\bar z})$
will in general also have a  nontrivial dependence on $\bar z$.
Quite analogously, the corresponding operator ${\bar t}(z, {\bar z})$,
which is related to ${\bar T}(\bar z)$ in the same way
as $t(z, {\bar z})$ is related to $T(z)$, can depend on $z$.

From the point of view of its $\bar z$ dependence, $t(z, {\bar z})$
must have dimension zero. Therefore, only two options are allowed for $t$.
 One is given by
$\bar L_0 t=0$.  This means
that either $t$ does not depend on ${\bar z}$,
or, that its ${\bar z}$-dependence
arises from a weight-zero operator, different from the identity.
The other option is given by
\begin{equation}
\bar L_0 t=T.
\end{equation}
This was first suggested in Ref.~\cite{Kogan2002}. See also
Ref.~\cite{Cardy2001}.

 In this note we focus entirely on the $z$-dependence.
The dependence of $t$ on $\bar z$ would have to treated
separately, and finally the holomorphic and antiholomorhpic
parts will have to be put together. However, we do
not address the issues of the ${\bar z}$-dependence
of $t(z, {\bar z})$  here. This is
an important problem which has not been properly investigated
 up until now, and should be a subject of future work.

\section{Extended Stress Tensor Multiplet}
\label{SectionExtendedStressTensorMultiplet}

In this section we first focus entirely on consequences of  conformal symmetry,
without regard to SUSY.  In particular, we will simply consider a general
CFT at $c=0$ which, as discussed in
Subsection \ref{SectionLogarithmicPartnerofStressTensor} above,
possesses a stress energy tensor $T$ as well as a logarithmic partner $t$,
satisfying the OPEs described in Eqs.~\rf{Tt} and \rf{txt}.  Such a theory
 may for example appear as   the $n\to 0$ limit of a replicated theory of $n$
interacting copies of fields, as discussed in the Introduction. We
are going to show that the requirement that $t(z)$ is logarithmic
automatically ensures the existence of extra primary  conformal
weight-two fields with nonvanishing two-point function. Moreover,
if there are  {\sl two} such extra fields,  $\xi(z)$ and $
\xi^\dagger(z)$, which are {\sl  anticommuting}, then the OPEs
between all four holomorphic weight-two fields $T, t, \xi,
\xi^\dagger$
 automatically transform covariantly under
a  global $U(1|1)$ supersymmetry.  In other words, we are going to show that
the requirements of conformal invariance, $c=0$, and the OPE
(\ref{Tt}), together then imply invariance under a global supersymmetry
acting on the holomorphic parts of these fields.
(We emphasize  that there was no mention of SUSY at the outset.)
Whether or not this  will in fact translate into
an actual global SUSY of the full $c=0$ theory will depend on the
specific gluing of holomorphic and antiholomorphic sectors,
an aspect that remains to be explored and which we are not addressing here.

\subsection{The multiplet of two dimensional operators, and SUSY}
\label{SubSectionTheMultipletOfTwoDimensionalOperatorsAndSusy}
So far, the operators with conformal weight two  are $t(z)$,
which is not primary according to the OPE \rf{Tt}, and $T(z)$, which
is primary at $c=0$, but has a vanishing two-point correlation
function.  It turns out, however, as we will now see,
that these two fields do not exhaust all conformal weight-two
operators we need to introduce for consistency of our theory.
We also need to add conformal weight-two primary operators
{\sl with nonvanishing correlation functions}.
% to the multiplet of such fields.

Consider, a  (chiral) correlation function with insertions of several
operators $t(z)$.
Such a  correlation function will not be single-valued.
In order to
%To
see how  the additional weight-two operators appear,
it is instructive to  examine the OPE
between two operators $t(z)$, given in \rfs{txt}.
Due to the presence of the logarithm,
this OPE is clearly not single-valued as $z$ goes around zero.
A correlation function with a  $t(z) t(0)$ insertion will change
under such a transformation.  A piece will be added to it whose
small $z$ behavior can be found by shifting all the
logarithms in \rf{txt} by $2 \pi i$.

It is possible to establish that this piece will contain a full
OPE between two primary operators $A$ with conformal weight two,
and nonzero
two-point correlation function. To be specific let us introduce the
conformal weight-two primary field $A_2 (z)$. Its OPE with itself
has the following contribution  from the identity operator
%goes as
\be
\label{AA2}
A_2(z) A_2(0) = \alpha T(z) T(0)+ {b \over 2 z^4}
+ { t(0)+  T(0) \log(z) \over z^2} + ... \ ,
\ee
in agreement with
\rf{AA}. Here we fixed the normalization of $A_2$ in a certain way
which will become useful later. We also introduced a piece in the
OPE proportional to the arbitrary coefficient $\alpha$. It is
allowed by conformal invariance and its utility will become
obvious as we go along. In the previous sections we routinely set
coefficients such as $\alpha$ to zero by employing the
redefinition \rf{redefine}, as well as the OPE (\ref{FactorizedTT})
of the stress tensor.  However, at this stage we
have already  used this redefinition once, to fix the OPE \rfs{txt}, so we cannot
use it again.  Notice that the three-point function \rf{tAA} is now
fixed to
\be
\label{tAA2}
\VEV{t(z) A_2(w_1) A_2(w_2)}=b\, {  \log
\Big( {w_1-w_2 \over (z-w_1) (z-w_2)} \Big)  + 2 \alpha \over
(z-w_1)^2 (z-z_2)^2 (w_1-w_2)^{2} } \,.
\ee
It follows that the OPE
$t(z) A_2(w)$ is fixed to be \be t(z) A_2(0) = 2 \alpha T(z)
A_2(0)- T(z) A_2(0) \log(z) + {A'_2(0) \over 2 z} +... \ ,
\ee
similar to the OPE \rf{tA}, but again with the
extra term proportional to $\alpha$. So far $\alpha$ remains arbitrary.

Now it can be checked, by  using (\ref{txt}) and (\ref{AA2}), that
the contribution of (the Virasoro representation of) the
identity operator in the following combination of OPEs
%\begin{eqnarray}
\begin{equation}
 \label{single}
t(z) t(0) %&+&
+4 A_2(z) A_2(0) \log(z) -T(z) T(0)
\ln^2(z)
%\nonumber\\[3mm]
%&+&
+\Big( {1\over 2}- 4 \alpha \Big)  T(z) T(0) \ln
(z)
\end{equation}
%\end{eqnarray}
no longer contains any logarithms, to all orders in $z$.
(A proof is presented in
Subsection \ref{SectionSubtractionLogs} of Appendix B.)
It therefore remains single-valued as a function of $z$, when inserted into any
correlation function.  This shows that indeed,
an insertion of $t(z) t(0)$ into any (chiral)
correlation function  with other operators, will analytically
continue to a linear combination of $t(z) t(0)$ and $A_2(z)
A_2(0)$ (and $T(z) T(0)$).  This is how the conformal weight-two primary
fields $A_2(z)$ will appear in any theory at $c=0$ which contains the logarithmic field
$t(z)$.

We already established in the Section
\ref{SubSectionOperatorsWithNonVanishingDimension} that (chiral)
correlation functions containing the conformal weight-two primary
operator $A_2(z)$
 (with nonvanishing two-point function)
would satisfy the corresponding
differential equation, if $b$ is either $5
\over 6$ or $- {5 \over 8}$. In the first case, this
would be the third order equation for the operator $A_{(3,1)}$, and in the
second the fifth order equation for the operator $A_{(1,5)}$.
In fact, the
solutions to these equations for the identity conformal block of the
four-point function
$\VEV{A_2 A_2 A_2 A_2}$ can be obtained
  in closed form.
An important feature of these solutions is that the identity conformal block
$F(x)$ defined as in \rf{ChiralFourPtFct}, \rf{DEFCrossRatio},
if normalized to $1$ as $x \rightarrow 0$, goes to  $(-x^4)$ as
$x \rightarrow \infty$, and vanishes  altogether when $x \rightarrow
1$ (or the opposite way around).
 This is only possible if  there  exist
{\it two} conformal weight-two operators
which are in fact {\sl anticommuting operators}.  Of course
we could also consider
{\it commuting} operators as far as the considerations leading to
Eq.~\rf{single} are concerned, but their correlation functions
would not satisfy the appropriate differential  equations at the
appropriate values of $b$.

Motivated by these considerations let us consider, from now on,
the case of two conjugate {\it fermionic} conformal weight-two  operators
$\xi(z)$ and $\xi^\dagger(z)$,
but we no longer require that they be Ka\v{c}-operators; in particular, the parameter $b$
can now take on arbitrary values.  (Besides the fermionic nature of $\xi$ and $\xi^\dagger$
we make no other assumptions but conformal symmetry.)
Their OPE can be copied from
\rf{AA2},
\be
\label{xx}
 \xi(z) \xi^\dagger(0) = \alpha T(z)
T(0)+ {b \over 2 z^4} + { t(0)+  T(0) \log(z) \over z^2} + ...\ .
\ee
It is remarkable that
the identity conformal block of the
four-point function of these operators,
\be
\label{GXiXidaggerXiXidagger}
G=\VEV{\xi(z_1) \xi^\dagger(z_2) \xi^\dagger(z_3) \xi(z_4) }\,,
\ee can
be obtained in closed form for arbitrary values of the parameter $b$ by
a  generalization of the conformal Ward identity. (Note that
unless $b=5/6$ or $b=-5/8$, this function no longer satisfies
a  corresponding differential equation.)
 In order to see this, consider
the  linear combination
\begin{eqnarray}
\VEV{\!\xi(z_1) \xi^\dagger(z_2) \xi^\dagger(z_3) \xi(z_4)\! }%&-&
-{1
\over 2} \!\VEV{\!T(z_1) T(z_2) \xi^\dagger(z_3) \xi(z_4)\!}
%\nonumber\\[3mm]
%&\times&
\log{
(z_1-z_2)(z_3-z_4) \over (z_1 - z_3) (z_2 -z_4)}
\,.\nonumber
\end{eqnarray}
 This
combination is a rational function of the coordinates, as can be
checked by employing the OPEs. Therefore, it can be reconstructed
%by
from  its poles, as in the standard conformal
Ward identity\cite{Belavin1984}.
This allows us to
find the correlation function $G$ for arbitrary values of $b$.
This correlation function must of course vanish as $z_1$
approaches $z_4$, due to the fermionic nature of the operators $\xi$.
It turns out that it
 vanishes only if the parameter $\alpha$ in (\ref{xx})
is chosen to be $\alpha={1 \over 8}$.
This explains why the parameter $\alpha$
was introduced  in (\ref{xx}) in the first place.
Setting $\alpha=1/8$, we obtain
%an
the explicit result,
\begin{eqnarray}
\label{GclosedFormSolution}
G&=&{b
\over {(z_1-z_2)^4 (z_3-z_4)^4}}\left[ { (x+1) (2 x^2 + b (x-1)^2
(1+x^2)) \over 4 (x-1) }\right. \nonumber\\[1mm]
& &\hskip 3.5cm \left. - \  {  x^2 (1-x+x^2)
\log(x) \over (x-1)^2} \right].
\end{eqnarray}
We can check by direct substitution that
(the function of the cross-ratio $x$, associated as in
\rf{ChiralFourPtFct} with the function)
$G$ satisfies the third
order equation for the operator $A_{(3,1)}$ if $b={5 \over 6}$, and
that it satisfies the fifth order equation for the operator $A_{(1,5)}$
if $b=- {5 \over 8}$.  It is not completely obvious from
(\ref{GclosedFormSolution}) that
$G \rightarrow 0$ when $z_1 \rightarrow z_4$ (which implies $x
\rightarrow 1$), but it can be easily checked with the help of
straightforward algebra.

Let us summarize briefly.
We have established that
$c=0$ conformal theories
with a logarithmic operator $t(z)$ must contain, in addition
to the stress tensor $T(z)$ and $t(z)$,
extra operators with conformal weight two but nonvanishing
two-point function.
For $b={5 \over 6}$ or $b=- {5 \over 8}$,
two {\sl anticommuting} operators $\xi$ and $\xi^\dagger$
are required, if the latter are to
satisfy the corresponding Ka\v{c}-table null-vector
conditions
(implying that the identity conformal block
of (\ref{GXiXidaggerXiXidagger}) satisfies the
3rd-, or the 5th-order differential equation,
corresponding to their respective Ka\v{c}-table positions).
In general, as long as these two conformal weight-two operator
$\xi$ and $\xi^\dagger$ are fermionic,
the full identity conformal
block of the function (\ref{GXiXidaggerXiXidagger}) can
be obtained in closed form
for any value of the `anomaly'
number $b$,
with the result given in
(\ref{GclosedFormSolution}) above.

These operators will then have the following OPEs
\be
\label{xxx} \xi(z)
\xi^\dagger(0) = {1 \over 8} T(z) T(0)+ {b \over 2 z^4} + { t(0)+
T(0) \log(z) \over z^2} + ...\,,
\ee
and
\be \label{tx} t(z)
\xi(0) = {1 \over 4} T(z) \xi(0)- T(z) \xi(0) \log(z) + {\xi'(0)
\over 2 z} + ...\,,
\ee
and similarly for  $\xi^\dagger$.

Consider now the OPEs given by Eqs.~\rf{FactorizedTT}, \rf{Tt},
\rf{txt}, \rf{xxx} and \rf{tx}. They were derived using CFT
techniques alone, and the only assumption was the existence of the
OPE \rf{Tt}, and the fermionic nature of the operators $\xi$ and
$\xi^\dagger$. Nevertheless, it can easily be verified that these
OPEs are in fact covariant under the application of a global SUSY
transformation according to Fig.~\ref{fig:mult}, acting only on
the chiral operators appearing in them. (Interestingly, these OPEs
would not have been covariant if $\alpha \not = {1 \over 8}$.)
Thus we arrive at a remarkable conclusion: if a partner of stress
energy tensor $t$ (which always exists at $c=0$ to avoid the $c
\rightarrow 0$ catastrophe) is logarithmic, these {\it chiral}
OPEs are automatically supersymmetric. This would have to be true
even when the field theory was constructed in terms of replicas as
opposed to supersymmetry, as long as there are {\sl two}
weight-two primary operators $\xi, \xi^\dagger$ with nonvanishing
two-point function  which are  {\sl fermionic}. (We know from
\rf{single} that there exists at least one such  weight-two
operator.)

\subsection{Comments on an `Extended Algebra'}
\label{SectionCommentsOnAnExtendedAlgebra}

 In Section \ref{LogarithmicAlgebra} we used the
logarithmic `algebra'
formed\,\footnote{\,%
`Algebra' appears here in quotes because the commutator  $[\ell_n,\ell_m]$
required by closure was not discussed; for an elaboration,
see the paragraphs below.}
by the coefficients of the mode
expansion of $T$ and $t$, which we called $L_n$ and $\ell_m$.
Specifically, we derived the commutation relation $[L_n,\ell_m]$, given by
\rf{CommuatorsLnln}.  It was enough to consider only this commutation
relation, besides the Virasoro algebra \rf{VirasoroAlgebra}, to arrive
at the results obtained in that Section.

It may be natural to ask if there could exist
a suitable  consistent (full)  extension of the Virasoro algebra, involving
the modes $\ell_m$.  For example, one could consider,
in addition to
% the
Virasoro descendants  $L_{-n}|A\rangle$,
also `extended Virasoro descendants' such as e.g. $\ell_{-n} |A\rangle$.
One may also ask about a possible generalization of null states,
degenerate with respect to a suitably extended Virasoro algebra,
 involving now also the modes $\ell_{+n}$.

Unfortunately, proposals along these lines have been difficult to implement.
In order to have a closed algebra, one would also need the commutation
relations $[\ell_n,\ell_m]$, which would have to be obtained
from the OPE \rf{txt}.  Similarly, since we now see that the full
set of weight-two fields involves not only
the fields $T$ and $t$, but also
additional fields such as e.g. the fermionic
fields  $\xi$ and $\xi^\dagger$, it seems natural at this point to try to establish
the anticommutation relations $\{\xi_n, \xi^\dagger_m \}$ as well.
One of the difficulties with these commutators is the nonholomorphic aspect of
these fields.  Furthermore,  regarding the second commutator
(but in view of (\ref{single}) this is also relevant for the first),
the operator $\xi$, when acting primary operators,
may generate other primary operators with
%may produce other primary operators with
different conformal weights.
%, when acting on primary operators.
For example, for the specific values of $b$ when the Ka\v{c} null-vector of  $\xi$
vanishes, its OPE with other primary fields can be read off
from  the Ka\v{c} table, and it obviously generates primaries with weights
not related by integers.  For arbitrary values of $b$ when the Ka\v{c} table is
not available, it is not even clear how to find such an OPE.
In any case, it would be interesting if a suitable extension of the Virasoro
algebra, based on the additional structures presented in this paper
could be developed, in one way or the other. This, however,  would certainly
have to be reserved for future work.\\[-8mm]~

\section{Conclusions}

In this paper we examined the structure of conformal field
theories (CFTs) with central charge $c=0$.
We focussed entirely on the holomorphic sector,
leaving the gluing of holomorphic and antiholomorphic
sectors
for future work.  One of the main
features distinguishing these CFTs from
those with $c\not = 0$ is the appearance of
a logarithmic `partner of the stress tensor'
which is in general not holomorphic.
It has a holomorphic part of conformal weight two,
which we denoted by $t(z)$
(Section \ref{SectionConformalFieldTheoryatczero}).
The latter leads to a novel anomaly number $b$,
distinguishing different CFTs with central charge
$c=0$. The number $b$ is a unique characteristic
of any given $c=0$ CFT.
 (Basically, $b$  plays the role of the
`effective central charge' $(\partial c(n)/\partial n)_{|n=0}$ of
replica theories \cite{Gurarie2002,Cardy2001}; recall the footnote
below \rf{corrTt}.) Interestingly, we saw in Section
\ref{LogarithmicAlgebra} that in theories with `logarithmic'
$t(z)$ the number $b$ controls the question of whether certain
Ka\v{c}-table null-vectors do indeed vanish identically or whether
they represent nonvanishing states with zero norm (like the $c=0$
stress tensor). We found that the null-vectors indeed vanish for
primary operators with nonvanishing two-point function  (i) in the
first two rows of the Ka\v{c} table when $b=+5/6$, or (ii)  in the
first column of the Ka\v{c} table with nonzero weight when
$b=-5/8$. Only those (chiral) four-point functions which contain a
Ka\v{c}-table operator with vanishing null-vector will satisfy the
familiar\cite{Belavin1984} differential equations.  These results
were obtained by considering
 suitably defined Laurent modes $\ell_{n}$ of the
logarithmic partner $t(z)$ of the stress tensor, and by
considering their commutation relations \rf{CommuatorsLnln} with
the ordinary Virasoro generators $L_n$. (This represents a
`partial extension' of the Virasoro algebra.) We showed in Section
\ref{SectionExtendedStressTensorMultiplet} that on grounds of
consistency there must exist, besides the stress tensor and its
partner $t(z)$, additional fields whose holomorphic parts have
weight two, and nonvanishing two-point functions, when $t(z)$ is
logarithmic. Remarkably, the simple assumption of a fermionic pair
of such additional fields implies that the  full set of chiral
OPEs between all these weight-two fields is automatically
covariant under the action of a global $U(1|1)$ supersymmetry. No
SUSY was required at the outset. (Then, also, the full identity
conformal block of these fermionic weight-two fields can be
computed exactly for any value of $b$, with the result given in
\rf{GclosedFormSolution}.) Indeed, a $U(1|1)$ multiplet of `stress
tensors' transforming in the same indecomposable representation
occurs in any CFT which is known to possess a  global $U(1|1)$
SUSY, the actual stress tensor being the singlet (see Section
\ref{SectionCriticalDisordered Systems}). Our results show that,
at least at the purely holomorphic level, such a global SUSY is
already a hidden symmetry in any $c=0$ CFT possessing a
logarithmic partner of the stress tensor $t(z)$, given there are
two fermionic weight-two fields $\xi, \xi^\dagger$ with
nonvanishing two-point function.  We
%end
close by saying that it is tempting to speculate
about a possible extension of the Virasoro algebra by the Laurent modes $\ell_{n}$ of the
logarithmic partner $t(z)$ of the stress tensor,
and by the corresponding modes of the other members of the
`stress tensor multiplet' mentioned above.
Future work will have to show if such an extension
can be constructed (Section \ref{SectionCommentsOnAnExtendedAlgebra}).

%\appendix

%\newpage

\section*{Appendix A:
Uniqueness of the `anomaly' number  \boldmath{$b$}}
%\addcontentsline{toc}{section}{Appendix A: Uniqueness of the `anomaly' number  \boldmath{$b$}}
\label{AppendixA}
\renewcommand{\theequation}{A.\arabic{equation}}
\setcounter{equation}{0}

In this appendix we will show that the OPE (\ref{Tt}),
which expresses the action of (infinitesimal)
conformal transformations
on the operator $t$, does not allow for two
{\it different} operators $t_1(z)$ and $t_2(z)$,
characterized by two {\it different} values of their
respective  `anomaly' numbers $b_1 \not =  b_2$,
to coexist in a given theory.

The proof is simple. If both,
$t_1(z)$ and $t_2(z)$,  were present in the same theory,
then we would be able to construct the (holomorphic)
two-point function $\langle t_1(z_1) t_2(z_2)\rangle $.
As in any CFT, this  function
must satisfy the constraints of global conformal invariance
(there are no others for a two-point function).
We will show that these constraints (ordinary differential
equations) do not possess a solution, unless $b_1=b_2$.

We start\,\footnote{\,We note in passing that the transformation
law under a finite conformal transformation $w=w(z)$ is
$
t(w) =({d z \over dw})^2 t(z) + [ \ln( {d z \over dw}) ]
\ T(z)$.}
by recalling that  the OPE (\ref{Tt}) yields
the change of the operators
$t_i(z)$, ($i=1,2$) under an infinitesimal
conformal transformation
$w(z)=z + \epsilon(z)$,
\begin{eqnarray}
\delta_{\epsilon(z)} \  t_i(z)
&=&\int_{C(z)} {  d \zeta \over 2 \pi i}
\ \epsilon(\zeta) \ T(\zeta)  t_i(z)
\nonumber\\[1mm]
\label{ConformalTransformationLaw}
&=&
\Big({d \over dz}
 \epsilon(z)\Big) [2 t(z) + T(z)]
+ \epsilon(z)
{d \over dz}
 t(z) +
{b_i\over 3!}\,
{d^3 \epsilon(z) \over dz^3}\ ,~~~
\end{eqnarray}
where  $i=1,\,\,2$. The action of the  global conformal group (Sl(2;{\bf C}))
corresponds to  functions $ \epsilon(z)$ which are 2nd order polynomials in $z$.
Now consider the two-point function
$$
\langle t_1(z_1) t_2(z_2)\rangle
$$
which is a function only of $z_{12}=z_1-z_2$,
due to translational invariance
(corresponding to $\epsilon(z)= constant$).
Choosing $\epsilon(z) = \epsilon \cdot z$ (scale invariance)
in (\ref{ConformalTransformationLaw})
yields
\begin{eqnarray}
&&\delta_{\epsilon(z)} \langle t_1(z_1) t_2(z_2)\rangle
=
 \langle\left (\delta_{\epsilon(z)} t_1(z_1)\right ) t_2(z_2)\rangle
+
 \langle t_1(z_1)\left (\delta_{\epsilon(z)}
 t_2(z_2)\right )\rangle
 \nonumber\\[1mm]
&&=
\left (\Big[ z_1{d \over dz_1 } + z_2{d \over dz_2 } \Big] +4\right )\langle t_1(z_1) t_2(z_2)\rangle
+ {b_1+b_2\over z_{12}^4}  =0\, ,
\end{eqnarray}
or (using translational invariance)
\be
z_{12} {d \over d z_{12}}
\left [  (z_{12})^4  \langle t_1(z_1) t_2(z_2)\rangle \right ]
+ (b_1+b_2) =0
\ee
%\be
%\nonumber
%\delta_{\epsilon(z)} \langle t_1(z_1) t_2(z_2)\rangle
%=
% \langle\left (\delta_{\epsilon(z)} t_1(z_1)\right ) t_2(z_2)>
%+
% \langle t_1(z_1)\left (\delta_{\epsilon(z)}
% t_2(z_2)\right )>
%\ee
%\be
%\! \!\!\!\! \!\!\!  \! \!\!\!\! \!\!\! \! \!\!\!\! \!\!\! \! \!\!\!\! \!\!\!  =
%\left (<[ z_1{d \over dz_1 } + z_2{d \over dz_2 } ] +4\right )<t_1(z_1) t_2(z_2)>
%+ {b_1+b_2\over z_{12}^4}  =0,
%\ee
%or (using translational invariance)
%\be
%z_{12} {d \over d z_{12}}
%\left [  (z_{12})^4<t_1(z_1) t_2(z_2)> \right ]
%+ (b_1+b_2) =0
%\ee
which has the solution
\be
\label{ScaleInvSolutions}
 (z_{12})^4 \langle t_1(z_1) t_2(z_2)\rangle
= - ( b_1+b_2) \ \ln(z_{12}) + {\rm const}.
\ee

Next, choosing $\epsilon(z) = \epsilon \cdot z^2$ (special
conformal transformations)
in (\ref{ConformalTransformationLaw}),
\begin{eqnarray}
%\! \! \! \! \!
&&\left\langle
\Big \{2 z_1 \left [ 2 t_1(z_1) + T(z_1)\right] + z_1^2 {d \over d z_1} t_1(z_1) \Big \}
 t_2(z_2) \right \rangle
\nonumber\\[1mm]
&&+
\left\langle t_2(z_2)
\Big\{2 z_2 \left[ 2 t_2(z_2) + T(z_2)\right] + z_2^2 {d \over d z_2} t_2(z_2)
\Big\}\right\rangle =0\,.
\end{eqnarray}
Setting $z_2=0$ yields
\be
z_1 {d \over d z_1} \left [
(z_1)^4 \langle t_1(z_1) t_2(z_2)\rangle
\right ]
+
2
\left [
(z_1)^4 \langle T(z_1) t_2(z_2)\rangle
\right ]
=0\,.
\ee
This gives, using
(\ref{Tt}) and (\ref{ScaleInvSolutions})
\be
-(b_1+b_2) + 2 b_2 =0,
\quad {\rm or} \ \ b_1=b_2\,.
\ee
This completes the proof.

Finally, we can read off from
(\ref{ScaleInvSolutions})
the two-point function of the single operator $t=t_1=t_2$,
\be
\label{ResultTwoPointtt}
\langle t(z_1) t(z_2)\rangle = {-2b \ln (z_{12}) + {\rm const}
\over z_{12}^4}\,,
\ee
as in (\ref{ttcor}) of the main text.

We close Appendix A by recalling that (\ref{ConformalTransformationLaw})
leads to the action of $L_n$ in the usual way\cite{Belavin1984},
by letting $\epsilon(z) \propto z^{n+1}$ ($n=-1, 0, 1, 2, ...$).
This yields, in particular, the relations in \rf{LnWitht} below.

%(\ref{LnAtwo}), (\ref{LnT}), (\ref{LnWitht})
%discussed

\section*{Appendix B:
Computation of OPEs of Virasoro descendants}
%\addcontentsline{toc}{section}{Appendix B: Computation of OPEs of
%Virasoro descendants}
\renewcommand{\theequation}{B.\arabic{equation}}
\setcounter{equation}{0}
\renewcommand{\thesubsection}{B.\arabic{subsection}}
\setcounter{subsection}{0}

This appendix is devoted to the computation of
OPEs such as for example (\ref{txt}).
More generally, consider instead of the two operators
$t(z)$ and $t(0)$ in that equation, two
%not necessarily primary
operators $O_1(z)$ and $O_2(0)$.
We are interested in the descendants
of some third operator of conformal weight $h'$,
appearing in the OPE $O_1(z)O_2(0)$.
%The operators $O_1(z)$ and $O_2(0)$ need not be primary.

Here we consider the OPE of the  two not necessarily primary operators
$ O_1(z)$ and $O_2(z)$ of conformal weights $h_1$ and $h_2$,
\be
\label{GeneralOPE}
 O_1(z)  O_2(0)
= ... + {1\over z^{h_1 + h_2 - h'}}\, X(z) + ...\,,
\ee
where the  ellipsis denotes contributions from
other primary operators.
(Examples are the OPEs in (\ref{AAOPE}), (\ref{AA}),
(\ref{AhoneAhtwoAhprimed}), (\ref{Tt}), (\ref{txt}).)
In general,
$X(z)$, which denotes the contributions
to this OPE from a primary operator of conformal
weight $h'$,
has the form
\be
\label{DefX}
X(z) = X^{(0)}(z) + [\ln(z)] X^{(1)}(z) + [\ln^2(z)] X^{(2)}(z)
+ ...\,,
\ee
where
$X^{(i)}(z)$, ($i=0, 1, 2, ...$)
are power series in $z$ whose coefficients
are operators evaluated at the point $z=0$,
denoting descendants (as well as their logarithmic `partners').
Explicitly, we have
\be
\label{XhatiExplicit}
X^{(i)}(z)
=
\sum_{n=0}^{\infty}  \ z^n \ {\hat X}^{(i)}_n(0) ,
\qquad \qquad (i=0, 1, 2, ...)\,,
\ee
where
$ {\hat X}_n^{(i)}(0) $ is
an operator of
conformal weight raised by  $+n$
as compared to $ {\hat X}_0^{(i)}(0) $,
or a  (certain, to-be-determined)
linear combination of
such operators.

Below, we will be interested in the logarithmic derivative
of $X(z)$.  Expanding $X(z)$ as in (\ref{DefX}) we have
\begin{eqnarray}
%\nonumber
\Big(z {d \over dz}\Big) X(z)
&\!=\!&
\left \{\! \Big(z {d \over dz}\Big) X^{(0)}(z)\! +\!  X^{(1)}(z) \!\right \}
%\\[3mm]
%&+&
\!+\!\ln(z)
\left \{\! \Big(z {d \over dz}\Big) X^{(1)}(z)\! +\! 2 X^{(2)}(z)\! \right \}
\nonumber\\[1mm]
&&+
\ln^2(z)
\left \{ \Big(z {d \over dz}\Big) X^{(2)}(z) + 3  X^{(3)}(z) \right \}
+ ...\,.
\label{RecursionXLogs}
\end{eqnarray}
Moreover, using the expansion (\ref{DefX}), (\ref{XhatiExplicit})
one obtains
\begin{equation}
%\begin{eqnarray}
%\nonumber
{1\over z^n} \left [ L_{+n}, \,X^{(i)} \right ]
%&=&
=
\sum_{m=n}^{\infty}  z^{m-n}
 \left [ L_{+n},  \,  {\hat X}^{(i)}_m(0) \right ]
% \\[3mm]
%&=&
=
\sum_{m=0}^{\infty}   z^{m}
 \left [ L_{+n},  \,  {\hat X}^{(i)}_{m+n}(0) \right ].
\label{PowerseriesLplusn}
%\end{eqnarray}
\end{equation}
Note that
$\left [ L_{+n},  \, {\hat X}^{(i)}_m(0) \right ] $
has conformal weight $(m-n)$ and that the
expression vanishes
when $n>m$.

Our aim in this appendix is to establish a recursion relation
to determine all the  coefficients of the entire power series, starting
from the first few (with lowest powers of $z$).
The fact that this is possible means that the entire OPE is
uniquely determined by its first few terms.

We start by considering the commutator of both
sides of (\ref{GeneralOPE}) with the
Virasoro (raising) operator $L_{+n}$, $n\geq 0$ (it
is actually enough to consider only $n=0, 1, 2$
because the others are generated using
the Virasoro algebra),
\be
\label{RecursionCommutator}
\left [ L_{+n},  \, O_1(z)  O_2(0) \right]
=
\left [
L_{+n},  \, X(z) \right]
+ {\rm {other \ prim.}}
\ee
The left-hand side  can we written as
\be
\label{RecursionCommutatorDerivation}
\left [ L_{+n},  \, O_1(z)  O_2(0) \right]
=
\left ( \left [ L_{+n},  \,  O_1(z) \right ] \right)
 O_2(0)
+
O_1(z)
\left ( \left [ L_{+n},  \, O_2(0) \right ] \right ).
\ee
These commutators are given by simple
expressions.

\vskip 0.5cm

\noindent{\it Examples of needed  commutators }

\vskip .3cm

\noindent We give  three examples of commutators
which we will need:

\vskip .3cm

\noindent (i)  for $A_h(z)$ a Virasoro primary of conformal weight $h$
\begin{eqnarray}
\label{LnAtwo}
n=0, 1, 2, ...:
\qquad
 \left [ L_{+n},  \, A_h(z) \right ]
&=&
z^n \left ( z {d \over dz} + h(n+1) \right ) A_h(z),
\qquad
\nonumber \\[1mm]
 \left [ L_{+n},  \, A_h(0) \right ]
&=& h \, \delta_{n,0} \, A_h(0);
\end{eqnarray}

\vskip .4cm

\noindent (ii)  for the stress tensor
\begin{eqnarray}
\hskip -1cm n=0, 1, 2, ...:
\qquad
\nonumber
 \left [ L_{+n},  \, T(z) \right ]
&=&
z^n \left ( z {d \over dz} + 2(n+1) \right ) T(z)
\nonumber\\[1mm]
&&+
{c\over 12} n(n^2-1) z^{n-2},
\qquad
\nonumber\\[2mm]
\label{LnT}
\left [ L_{+n},  \ T(0) \right ]
&=&2 \delta_{n,0} \, T(0) +
 \delta_{n,2} \   {c\over 2}\ ;
\end{eqnarray}

\vskip .3cm

\noindent (iii)  for the `partner' $t(z)$ of the stress tensor
\begin{eqnarray}
\nonumber
n=0, 1, 2, ...:
%&&
%\qquad
%\qquad
%\qquad
%\qquad
%\qquad
%\\[2mm]
%\nonumber
\quad
 \left [ L_{+n},  \, t(z) \right ]
&=&
z^n
\left \{\!
 \left (\! z {d \over dz} \!+\! 2(n+1) \!\right ) t(z)\! +\!
(n+1) T(z) \right \}
\\[1mm]
\nonumber
&&+
{b\over 6} n(n^2-1) z^{n-2}\,,
\\[2mm]
\label{LnWitht}
 \left [ L_{+n},  \, t(0) \right ]
&=&
\delta_{n,0}
\biggl (2 t(0) + T(0)\biggr ) +
 \delta_{n,2} \ b\,.
\end{eqnarray}

\subsection{The OPE of two primary operators,
$A_{h_1}(z)A_{h_2}(0)$ }
\label{AppBTheOPEofTwoPrimaryOperators}

We now proceed to derive the recursion relations for the OPE of
two primary
operators of conformal weights $h_1$ and $h_2$,
respectively,
\be
\label{OoneOtwoAtwo} O_1=A_{h_1}, \qquad
O_1=A_{h_2}\,.
\ee
We are interested in the descendants of a third
primary operator $A_{h'}$
of conformal weight $h'$ appearing in this OPE, which we
characterize by the (operator-valued) function $X(z)$,
\begin{eqnarray}
&&A_{h_1}(z)A_{h_2}(0) = ... + {1\over z^{h_1 + h_2 - h'}} X(z) +
...
\nonumber\\[1mm]
\label{AhoneAhtwoAhprimed}
&&=... + {1\over z^{h_1 +
h_2 - h'}} \left \{ A_{h'}(0) + a_1 \ z \
 L_{-1} A_{h'}(0)
+ O(z^2)
\right \}
+ ...\,,
\end{eqnarray}
where the  ellipsis denote contributions from other primary operators.
Making use of
(\ref{RecursionCommutator}), (\ref{RecursionCommutatorDerivation}), (\ref{LnAtwo})
for $n\geq 1$ yields
\be
z^n \left [
\Big(z{d \over dz}\Big) + h_1 (n+1)
\right ]
A_{h_1}(z) A_{h_2}(0)
=
... + {1\over z^{h_1 + h_2 - h'}}
\left [
L_{+n}, X(z)
\right ]
+ ...
\ee
which becomes\,\footnote{\,We used
$ z^x \left ( z {d \over dz} \right )  \left ( z^{-x} X(z) \right )=$
$\left ( z {d \over dz} - x \right ) X(z)$.}
\be
\label{CommutatorLnforAtwo}
n \geq 1:
\quad
%\left \{
\left ( z {d \over dz} +
 n h_1 -h_2 + h'
\right )
%\right \}
 X(z)
=
{1\over z^n}  \left [ L_{+n},  \, X(z) \right ].
\ee
Using (\ref{RecursionXLogs}) and (\ref{PowerseriesLplusn})
in (\ref{CommutatorLnforAtwo})
we obtain  a recursion for the coefficients
\begin{eqnarray}
\nonumber
\!  \!\!\!\! \!\!\!\!
 n \geq 1: \quad
%\left \{
 [m+n h_1 - h_2 + h']
{\hat X}^{(0)}_m + {\hat X}^{(1)}_m
%\right \}
&=&
 \left [ L_{+n},  \, {\hat X}^{(0)}_{m+n}(0) \right ],  \\[1mm]
\nonumber
%\left \{
 [m+ n h_1 - h_2 + h']
 {\hat X}^{(1)}_m + 2 {\hat X}^{(2)}_m
%\right \}
&=&
 \left [ L_{+n},  \, {\hat X}^{(1)}_{m+n}(0) \right ],  \\[1mm]
%\left \{
 [m+ n h_1 - h_2 + h']
{\hat X}^{(2)}_m + 3 {\hat X}^{(3)}_m
%\right \}
&=&
 \left [ L_{+n},  \ {\hat X}^{(2)}_{m+n}(0) \right ] ,
\label{RecursionExplicitAtwoAtwo}
\end{eqnarray}
etc.. Upon choosing $n=1,2$,
these equations determine the coefficients
$ {\hat X}^{(0)}_{m}(0), {\hat X}^{(1)}_{m}(0), {\hat X}^{(2)}_{m}(0)$
for higher values of the index from those with a lower index.

\vskip .1cm

For $n=0$ (dilations) we have
\be
\left [
L_0, A_{h_1}(z) A_{h_2}(0)
\right ]
=
\left(\Big(
z {d \over dz}\Big) + (h_1+h_2)
\right)
A_{h_1}(z) A_{h_2}(0),
\ee
or
\be
\left(
\Big(z {d \over dz}\Big) +  h'
\right)
X(z)
=
\left [
L_0,  X(z)
\right ].
\ee
This yields for the coefficients of $X(z)$
\begin{eqnarray}
\nonumber
%\left \{
(m+ h')
{\hat X}^{(0)}_m + {\hat X}^{(1)}_m
%\right \}
&=&
 \left [ L_{0},  \, {\hat X}^{(0)}_{m} \right ] , \\[1mm]
\nonumber
%\left \{
(m+ h')
 {\hat X}^{(1)}_m + 2 {\hat X}^{(2)}_m
%\right \}
&=&
 \left [ L_{0},  \, {\hat X}^{(1)}_{m} \right ]  ,\\[1mm]
%\left \{
(m+  h')
{\hat X}^{(2)}_m + 3 {\hat X}^{(3)}_m
%\right \}
&=&
 \left [ L_{0},  \ {\hat X}^{(2)}_{m} \right ] ,
\label{RecursionExplicitAoneAtwoLzero}
\end{eqnarray}
etc..

\subsection{The OPE $T(z) T(0) \sim {\bf 1}$ at $c=0$ }

In order to compute this OPE we set
\be
\label{TandT}
O_1=O_2 =T\,
\ee
and $h'=0$ in \rf{GeneralOPE}. Making use of
(\ref{RecursionCommutator}), (\ref{RecursionCommutatorDerivation}), (\ref{LnT})
for $n \geq 1$ we obtain
\be
%\nonumber
n \geq 1: \qquad
z^n
\left \{
\left ( z {d \over dz} + 2(n+1) \right )T(z)\ T(0)
\right \}
=
{1\over z^4}  \left [ L_{+n},  \, X \right ].
\quad
%\qquad
\ee
This has the same form as the equation (\ref{CommutatorLnforAtwo})
obtained above, for two  conformal weight $=2$ primary operators $A_2$
(i.e. $h_1=h_2=2$).
This is of course expected, because the stress tensor $T(z)$
is weight-two primary, at $c=0$.
These two  types of primary weight-two operators
differ only by the fact that one
(the stress tensor)
has a vanishing, and the other ($A_2$) a nonvanishing
two-point function.  The recursion relations are thus identical to
(\ref{RecursionExplicitAtwoAtwo}), with $h_1=h_2=2$.
But the terms in $X(z)$ with small powers of $z$,
  i.e. the initial conditions of the recursion
relations, are different in the two cases.
Since for the stress tensor $T(z)$
the initial conditions of the recursion do not
contain any logarithms,
this continues to be case for all terms,
and the recursion relations
(\ref{RecursionExplicitAtwoAtwo}) simplify further,
\begin{eqnarray}
\label{RecursionExplicitTwithTSimplified}
n \geq 1: \qquad
%\left \{
 [m + 2(n-1)] {\hat X}^{(0)}_m
%\right \}
&=&
 \left [ L_{+n},  \, {\hat X}^{(0)}_{m+n}(0) \right ].
\end{eqnarray}

\subsection{The OPE $t(z) A_h(0) \sim A_h(0)$}

In this section of Appendix B we determine the terms appearing
in the OPE \rf{tA} of
Subsection \ref{SubSectionOperatorsWithNonVanishingDimension}.
We write this OPE in the form
\be
\label{OPEtwithA}
t(z) A_h(0)
:=
{1\over z^2}\, X(z) + \Re(z)\,,
\ee
where $\Re(z)$ denotes all those terms in this OPE
which contain noninteger powers of $z$.

\vskip .2cm

\noindent {(i)} \ \
Applying
\rf{RecursionCommutator}, \rf{RecursionCommutatorDerivation}
to
\rf{OPEtwithA}
yields
\begin{eqnarray}
\nonumber
&&n \geq 1:
%\qquad
%\qquad \qquad \qquad \qquad
%\qquad \qquad \qquad \qquad
%\qquad \qquad \qquad \qquad
\\[2mm]
\nonumber
&&z^n
\left \{
\left (
z {d \over dz} + 2 (n+1) \right ) t(z)
+
(n+1) T(z)
+
{b\over 6}\, n (m^2-1) z^{-2}
\right \}
A_h(0)
\nonumber\\[1mm]
&&=
\left [ L_n,
{1\over z^2}\, X(z) + \Re(z)
\right ].
\label{OPEtwithAOne}
\end{eqnarray}
This becomes, abbreviating
\be
\label{DEFYofZforTA}
T(z) A_h(0) := {1\over z^2}\, Y(z),
\ee
\begin{eqnarray}
\nonumber
&&n \geq 1:
%\qquad
%\qquad \qquad \qquad \qquad
%\qquad \qquad \qquad \qquad
%\qquad \qquad \qquad \qquad
\\[2mm]
\nonumber
&&
z^2\left (
z {d \over dz} + 2(n+1) \right )
z^{-2} X(z)
+ (n+1) Y(z)
+ {b\over 6} n(n^2-1) A(0)
\\[1mm]
&&=
{1\over z^n} \left [
L_n, X(z) \right ] .
\label{OPEtwithATwo}
\end{eqnarray}
Using the footnote in Subsection
\ref{AppBTheOPEofTwoPrimaryOperators}
the above  reduces to
\begin{eqnarray}
\nonumber
&&n \geq 1:
%\qquad
%\qquad \qquad \qquad \qquad
%\qquad \qquad \qquad \qquad
%\qquad \qquad \qquad \qquad
\\[2mm]
\nonumber
&&
\left (
z {d \over dz} +2 n \right )
X(z)
+ (n+1) Y(z)
+ {b\over 6} \,n(n^2-1) A(0)
\\[1mm]
&&=
{1\over z^n} \left [
L_n, X(z) \right ] .
\label{OPEtwithAFinal}
\end{eqnarray}

%\vskip .2cm

\noindent {(ii)} \ \
Similarly, for $n=0$ we obtain from
\rf{RecursionCommutator}, \rf{RecursionCommutatorDerivation}
and
\rf{OPEtwithA}
\begin{eqnarray}
\nonumber
&&n = 0:
%\qquad
%\qquad \qquad \qquad \qquad
%\qquad \qquad \qquad \qquad
%\qquad \qquad \qquad \qquad
\\[2mm]
%\nonumber
&&
\left (
z {d \over dz} +2 +h \right )
t(z) A_h(0)
+T(z) A_h(0)
%\\[3mm]
%&&=
=\left [
L_0, {1\over z^2} X(z) + \Re(z)
\right ] ,~~~~~~~
\label{OPEtwithAnzero}
\end{eqnarray}
leading to
\begin{eqnarray}
%\nonumber
&&n = 0:
\qquad
%\qquad \qquad \qquad \qquad
%\qquad \qquad \qquad \qquad
%\qquad \qquad \qquad \qquad\\[2mm]
%&&
\left (
z {d \over dz}  +h \right )
X(z)
+ Y(z)
=
\left [
L_0,  X(z)
\right ] .
\label{OPEtwithAnzeroFinal}
\end{eqnarray}

\noindent {(iii)}  \ \ \
Finally, inserting the decomposition \rf{DefX},
using \rf{RecursionXLogs} as well as \rf{PowerseriesLplusn},
and recalling that the quantity $Y(z)$
defined in \rf{DEFYofZforTA} has no terms proportional to $\ln(z)$,
we obtain

\vskip .1cm

\noindent $*$ \  from \rf{OPEtwithAFinal},
\begin{eqnarray}
\nonumber
n \geq 1: \qquad
&&\left \{
 [m+ 2 n ]
{\hat X}^{(0)}_m +
(n+1){\hat Y}^{(0)}_m+ {b\over 6}\,n(n^2-1) A(0)+
 {\hat X}^{(1)}_m
\right \}
\\[1mm]
\nonumber
&&=
 \left [ L_{+n},  \ {\hat X}^{(0)}_{m+n}(0) \right ] , \\[2mm]
&&\left \{
 [m+  2n ]
 {\hat X}^{(1)}_m + 2 {\hat X}^{(2)}_m
\right \}
=
 \left [ L_{+n},  \ {\hat X}^{(1)}_{m+n}(0) \right ]  ,
\label{RecursionExplicittwithA}
\end{eqnarray}
etc.

\vskip 2mm

\noindent $*$ \  from \rf{RecursionExplicitAoneAtwoLzero},
\begin{eqnarray}
\nonumber
n=  0:
\qquad
&&\left \{
(m+ h)
{\hat X}^{(0)}_m +
{\hat Y}^{(0)}_m
+ {\hat X}^{(1)}_m
\right \}
=
 \left [ L_{0},  \ {\hat X}^{(0)}_{m} \right ] , \\[1mm]
\nonumber
&&\left \{
(m+ h)
 {\hat X}^{(1)}_m + 2 {\hat X}^{(2)}_m
\right \}
=
 \left [ L_{0},  \ {\hat X}^{(1)}_{m} \right ] ,
\label{RecursionExplicittwithALzero}
\end{eqnarray}
etc..

\subsection{The OPE $t(z) t(0)\sim {\bf 1}$, \ ($c=0$)}
\label{TheOPEgztzero}

In order to compute this OPE we set
\be
\label{TandT1} O_1=O_2 =t\,
\ee
and $h'=0$ in \rf{GeneralOPE}.
Making use of (\ref{RecursionCommutator}),
(\ref{RecursionCommutatorDerivation}), (\ref{LnWitht}) for $n \geq
1$
\begin{eqnarray}
\nonumber
&&\left \{ \left[ (z {d \over dz} + 2(n+1) \right
]t(z) +(n+1) T(z) + {b\over 6} \,n(n^2-1) z^{-2} \right \} t(0)
\\[1mm]
&&+
t(z) b \delta_{n,2} z^{-2}
={1\over z^4} \ {1\over z^n}
[L_{+n}, X] + ...\, , \quad (n \geq 1).
\end{eqnarray}
We now use the definition
\be
\label{DefYzInhomogeneity} Y(z) := \{ b + z^2 \left ( 2t(0) + T(0)
\right ) + z^3 L_{-1}t(0) + ... \}
\ee
for the term arising from
the OPE $T(z)t(0)=Y(z)/z^4$, discussed in (\ref{Tt}). Note that the
so-defined function $Y(z)$ has an analytic $z$-dependence, and hence no
logarithms. We denote this fact by writing (in view of the
notation used in (\ref{DefX}))
 \be
 \label{YnoLogs}
 Y(z) =
Y^{(0)}(z) \,.
\ee
Thus, we may write this as
\begin{eqnarray}
\nonumber
&&\left[ (z
{d \over dz} + 2(n-1) \right ] X(z) +(n+1) Y^{(0)}(z)
\\[1mm]
\label{PreviousEquationtt}
&&+ z^2 \left
[ {b\over 6}\, n(n^2-1) t(0) + b \delta_{n,2} t(z) \right ] =
   {1\over z^n}
[L_{+n}, X(z)],
\end{eqnarray}
at $n \geq 1$.
To simplify the notation,
we also define
\be
\label{DefZtandt}
Z(n;z)
:=
z^2 \left [ {b\over 6} n(n^2-1)  t(0) + b \delta_{n,2} t(z) \right ],
\ee
and
\be
Z(0;z)
=
Z(1;z)=0,
\ee
Moreover,
\be
Z(2;z)=
z^2  b\left [  t(0) +  t(z) \right ],
\quad
Z(n;z)
:=
z^2 \, {b\over 6} \,n(n^2-1)  t(0),
\ee
 for $n\geq 3$, where
\be
t(z)
=
\sum_{n=0}^\infty {z^n\over n!} \Big({d \over dz}\Big)^n t(0),
\ee
in parallel to
\begin{eqnarray}
\nonumber
T(z) &=& T(z) {\bf 1}(0)
=
\sum_{n=0}^\infty \ z^{n} L_{-2-n}{\bf 1}(0)=
\sum_{n=0}^\infty {z^n\over n!} \Big({d \over dz}\Big)^n T(0),\\[1mm]
\label{ExpansionT}
T(0)&=&L_{-2}{\bf 1}(0).
\end{eqnarray}
Hence, we see that
\be
Z(n;z)
=
Z^{(0)}(n;z)
\ee
contains no logarithms.

Now, we may write  the previous equation, \rf{PreviousEquationtt},
in the following final form
(recalling the definitions (\ref{DefYzInhomogeneity}) and
(\ref{DefZtandt}))
\be
\nonumber
\left[
\Big(z {d \over dz}\Big) + 2(n-1)
\right ] X(z)
+(n+1)
Y^{(0)}(z)
+
Z^{(0)}(z)
=
   {1\over z^n}\,
[L_{+n}, X],
\quad (n \geq 1).
\ee
The recursion for the coefficients now reads
\begin{eqnarray}
\nonumber
&& n \geq 1:
%\qquad
%\qquad
%\qquad
%\qquad
%\qquad
%\qquad
%\qquad
\\[1mm]
\nonumber
 && [m + 2(n-1)] {\hat X}^{(0)}_m + {\hat X}^{(1)}_m
+(n+1) {\hat Y}^{(0)}_m +{\hat Z}^{(0)}_m(n)
=
 \left [ L_{+n},  \ {\hat X}^{(0)}_{m+n}(0) \right ] , \\[1mm]
\nonumber
&&
 [m + 2(n-1)]
 {\hat X}^{(1)}_m + 2 {\hat X}^{(2)}_m
=
 \left [ L_{+n},  \ {\hat X}^{(1)}_{m+n}(0) \right ] , \\[1mm]
&&
 [m + 2(n-1)]
{\hat X}^{(2)}_m
=
 \left [ L_{+n},  \ {\hat X}^{(2)}_{m+n}(0) \right ] .
\label{RecursionExplicitTandT}
\end{eqnarray}
(We have omitted terms containing a triple power of  the logarithm,
as they will not be generated.)

\vskip .1cm

For $n=0$ we obtain from
(\ref{RecursionCommutator}), (\ref{RecursionCommutatorDerivation}), (\ref{LnWitht})
\be
\left [
L_0, t(z) t(0)
\right ]
=
\left \{
\left (
\Big(z {d \over dz}\Big) +2
\right )
t(z)
+
T(z)
\right \} t(0)
+
t(z) \bigl ( 2 t(0) + T(0)
\bigr )
\quad
\ee
or
\be
\Big(z{d \over dz}\Big) X(z) + Y^{(0)}(z) +
z^4 t(z) T(z) = \left [
L_0, X(z)
\right ],
\ee
leading to
\begin{eqnarray}
\nonumber
%\left \{
m {\hat X}^{(0)}_m +
 {\hat X}^{(1)}_m +
 {\hat Y}^{(0)}_m + z^4 t(z) T(z)
%\right \}
&=&
 \left [ L_{0},  \, {\hat X}^{(0)}_{m}(0) \right ],
 \\[1mm]
\nonumber
%\left \{
 m
 {\hat X}^{(1)}_m + 2 {\hat X}^{(2)}_m
%\right \}
&=&
 \left [ L_{0},  \, {\hat X}^{(1)}_{m}(0) \right ],
 \\[1mm]
\label{RecursionExplicitTandTLzero}
%\left \{
 m
{\hat X}^{(2)}_m
%\right \}
&=&
 \left [ L_{0},  \, {\hat X}^{(2)}_{m}(0) \right ].
\end{eqnarray}

\vskip 1cm

\subsection{Summary of Recursions}

\vskip .5cm
\underbar{\it Notation:}
\begin{eqnarray}
X(z) &=& X^{(0)}(z) + [\ln(z)] X^{(1)}(z) + [\ln^2(z)] X^{(2)}(z)
+...\,.
\nonumber\\[1mm]
X^{(i)}(z)
&=&
\sum_{n=0}^{\infty}  \ z^n \ {\hat X}^{(i)}_n(0) ,
\quad \quad (i=0, 1, 2, ...).
\end{eqnarray}
(We omit the position $(0)$ in the formulas below.)

%\vskip 1cm
%\newpage

\be
\bullet \ \
A_h(z) A_h(0) ={1\over z^{2h}} X^{(A)}(z) + ...\, ,
\quad
\left ( {\rm identity \ operator,} \ \ h'=0
\right)\,,
\ee
\begin{eqnarray}
\nonumber
&& m=0, 1, 2, ...
%\qquad
%\qquad
%\qquad
%\qquad
\\[1mm]
\nonumber
&&
n \geq 1: \   \ \
 [m + h(n-1)] {\hat X}^{(A;0)}_m + {\hat X}^{(A;1)}_m
=
 \left [ L_{+n},  \, {\hat X}^{(A;0)}_{m+n} \right ],
 \\[1mm]
\label{RecursionExplicitAtwoAtwoSimplifiedSummary}
&&\qquad\qquad ~[m + h(n-1)]
{\hat  X}^{(A;1)}_m
=
 \left [ L_{+n},  \, {\hat X}^{(A;1)}_{m+n} \right ],
 \\[3mm]
\nonumber
&& n=0: \ \
\qquad
m {\hat X}^{(0)}_m + {\hat X}^{(1)}_m
= \left [ L_{0},  \, {\hat X}^{(0)}_{m} \right ],
 \\[1mm]
\label{RecursionExplicitAoneAtwoLzeroSummary}
&&\qquad\qquad\qquad m {\hat X}^{(1)}_m
=
 \left [ L_{0},  \, {\hat X}^{(1)}_{m} \right ].
\end{eqnarray}

%\vspace{.3cm}

$$
\bullet \  \
T(z) T(0) ={1\over z^4} \,X^{(T)}(z)\,,
\qquad
\qquad
\qquad
\qquad
\qquad
\qquad
\qquad
\qquad
\qquad
$$
\begin{eqnarray}
&&n \geq 1, \quad  m=0, 1, 2, ...:
\nonumber\\[2mm]
&&
 [m + 2(n-1)] {\hat  X}^{(T;0)}_m =
 \left [ L_{+n},  \ {\hat X}^{(T;0)}_{m+n} \right ]
.
\label{RecursionExplicitTwithTSimplifiedSummary}
\end{eqnarray}

\vskip .5cm

\noindent \quad \quad
$\bullet$ \ $t(z) A_h(0) = {1\over z^{2}} \,X^{(tA)}(z)+ \Re(z)$,
 \ \  $T(z) A_h(0) = {1\over z^{2}} Y(z)$,
\qquad
\quad
\quad
\begin{eqnarray}
\nonumber
&&m=0, 1, 2, ...
\qquad \qquad \qquad \qquad
\qquad \qquad \qquad \qquad
\qquad \qquad
\\[1mm]
\nonumber
&& n \geq 1: \quad
 [m+ 2 n ]
{\hat X}^{(tA;0)}_m +
(n+1) {\hat Y}^{(0)}_m
\\[3mm]
\nonumber
&&\qquad\qquad+ {b\over 6}\,n(n^2-1) A(0)+
 {\hat X}^{(tA;1)}_m
=
 \left [ L_{+n},  \ {\hat X}^{(tA;0)}_{m+n} \right ],  \\[1mm]
&&
\qquad\qquad [m+  2n ]
 {\hat X}^{(tA;1)}_m + 2 {\hat X}^{(tA;2)}_m
=
 \left [ L_{+n},  \ {\hat X}^{(tA;1)}_{m+n} \right ] ,
\label{RecursionExplicittwithASummary}
\end{eqnarray}
etc.
\begin{eqnarray}
\nonumber
&&n=  0:
\qquad
\nonumber
(m+ h)
{\hat X}^{(tA;0)}_m +
{\hat Y}^{(0)}_m + {\hat X}^{(tA;1)}_m
=
 \left [ L_{0},  \ {\hat X}^{(tA;0)}_{m} \right ],  \\[1mm]
&&\qquad\qquad\quad
(m+ h)
 {\hat X}^{(tA;1)}_m + 2 {\hat X}^{(tA;2)}_m
=
 \left [ L_{0},  \ {\hat X}^{(tA;1)}_{m} \right ] ,
\label{RecursionExplicittwithALzeroSummary}
\end{eqnarray}
etc.
(Note that the combination
$ {\hat Y}^{(0)}_m + {\hat X}^{(tA;1)}_m$
appearing in the first of these two equations
must vanish, because ${\hat X}^{(tA;0)}_{m}$
has weight $(h+m)$.  This  determines
$ {\hat X}^{(tA;1)}_m$ with the same result as
in  Section \ref{SectionDetailsOPEtwithA} below.)
\be
\nonumber
\bullet \  \  \
t(z) t(0) ={1\over z^4} X^{(t)}(z)+...\, ,
\quad
T(z) t(0) ={1\over z^4} Y^{(t)}(z),
\qquad
\qquad
\qquad
\qquad
\qquad
\qquad
\qquad
\qquad
\ee
\begin{eqnarray}
\nonumber
&&n \geq 1;  \  m=0, 1, 2, ...:
\qquad
\qquad
\qquad
\qquad
\qquad
\qquad
\\[1mm]
\nonumber
&&
 [m + 2(n-1)] {\hat X}^{(t;0)}_m + {\hat X}^{(t;1)}_m
+(n+1) {\hat Y}^{(t;0)}_m +{\hat Z}^{(t;0)}_m(n)
=
 \left [ L_{+n},  \ {\hat X}^{(t;0)}_{m+n} \right ] , \\[1mm]
\nonumber
&& [m + 2(n-1)]
 {\hat X}^{(t;1)}_m + 2 {\hat X}^{(t;2)}_m
=
 \left [ L_{+n},  \ {\hat X}^{(t;1)}_{m+n} \right ] , \\[1mm]
&&
 [m + 2(n-1)]
{\hat X}^{(t;2)}_m
=
 \left [ L_{+n},  \ {\hat X}^{(t;2)}_{m+n} \right ] ,
\label{RecursionExplicitTandTSummary}
\end{eqnarray}
\vspace{1mm}
\begin{eqnarray}
\nonumber
&&n=0:
\quad
m {\hat X}^{(t;0)}_m +
 {\hat X}^{(t;1)}_m +
 {\hat Y}^{(0)}_m + z^4 t(z) T(z)
=
 \left [ L_{0},  \ {\hat X}^{(t;0)}_{m} \right ],
 \\[1mm]
\nonumber
&&\qquad\qquad
 m
 {\hat X}^{(t;1)}_m + 2 {\hat X}^{(t;2)}_m
=
 \left [ L_{0},  \ {\hat X}^{(t;1)}_{m} \right ],
 \\[1mm]
\label{RecursionExplicitTandTLzero1}
&&\qquad\qquad\left \{
 m
{\hat X}^{(t;2)}_m
\right \}
=
 \left [ L_{0},  \ {\hat X}^{(t;2)}_{m} \right ].
\end{eqnarray}
(Again, we have omitted terms containing a triple power of  the logarithm,
as they will not be generated.)

\subsection{Details of OPE $A_h(z) A_h(z) \sim {\bf 1}$}
 \label{SectionDetailsOPEAtwoAtwo}

The form of the descendants of the identity operator appearing in the
OPE $A_h(z) A_h(z)$ of two primary operators, in the normalization of (\ref{AA}),
corresponds to the  coefficients
\begin{eqnarray}
\nonumber
{X}^{(A;0)} &=&
 1
+ z^2\, {h\over b}\, t(0)
 + O(z^3),
\\[1mm]
\label{AandAXofz}
{X}^{(A;1)} &=&
 \quad
 \quad
z^2 \,{h \over b}\, T(0) +O(z^3)\,.
\end{eqnarray}
Let us illustrate how the order $O(z^2)$ terms, and similarly all the others,
are obtained from the leading term in (\ref{AandAXofz}) by applying the recursion.

\noindent The lowest order terms  of (\ref{AandAXofz})  read
%lead to
\be
{X}^{(A;0)}_0=1;
\qquad
{X}^{(A;1)}_0=0\,.
\ee
Using
(\ref{RecursionExplicitAtwoAtwoSimplifiedSummary})
with $m=0$ and $n=2$  leads to
%reads
\begin{eqnarray}
\label{DetailsOPEAhAh}
h + 0 &=&  \left [ L_{+2}, {X}^{(A;0)}_2 \right]\,,
\\[1mm]
0 &=& \left [ L_{+2}, {X}^{(A;1)}_2 \right]\,,
\end{eqnarray}
which implies, using (\ref{LnT}), (\ref{LnWitht})
\begin{eqnarray}
\nonumber
{X}^{(A;0)}_2 &=&  {h\over b} \,\big( t(0) + \alpha  \ T(0) \big)\,,
\\[1mm]
{X}^{(A;1)}_2 &=&  {h\over b} \,\beta  \, T(0)\,.
\end{eqnarray}
Using
(\ref{RecursionExplicitAoneAtwoLzeroSummary})
with $m=2$ we get
\begin{eqnarray}
\nonumber
2{X}^{(A;0)}_2
+{X}^{(A;1)}_2
 &=& \left [ L_{+0}, {X}^{(A;0)}_2 \right ]\,,
\\[1mm]
2{X}^{(A;1)}_2 &=& \left [ L_{+0}, {X}^{(A;1)}_2\right ]\,,
\end{eqnarray}
which yields, when setting the arbitrary constant
$\alpha \to 0$
\begin{eqnarray}
\nonumber
2 t(0)+ \beta  \ T(0) &=& 2t(0) + T(0)\,,
\\[1mm]
2 \beta  \ T(0)
  &=&  2 \beta  \, T(0)\,.
\end{eqnarray}
Hence we have found  $\beta =1$ ($\alpha=0$), in
agreement with (\ref{AandAXofz}).
(Note that $\alpha$ is arbitrary because if corresponds
(at $c=0$) to the contribution of a primary operator $T(z)$
(the stress tensor) to the OPE $A_h(z) A_h(z)$.)

\subsection{Details of OPE $t(z) A_h(z) \sim A_h(z) $}
\label{SectionDetailsOPEtwithA}

We begin by writing down the leading terms in the OPEs
\rf{DefX}, \rf{OPEtwithA}, \rf{DEFYofZforTA},
\begin{eqnarray}
\nonumber
{X}^{(tA;0)} &=& 0 \qquad  \quad   + O(z)\,,
\\[1mm]
\nonumber
{X}^{(tA;1)}
&=&- h A(0) +  O(z)\,,
\\[1mm]
\label{tandAXofz}
Y(z)
&=& h A(0)  \ \  + z L_{-1}A(0)+ O(z^2)\,.
\end{eqnarray}
Recall that in the notation of \rf{tA}
\be
\label{NotationXandtA}
{X}^{(tA;0)}
=
z \  \ell_{-1} A(0)
+
z^2  \ \ell_{-2} A(0) + ...\,.
\ee

\noindent {\it a}) \ \ We start with
\rf{RecursionExplicittwithASummary}
for $m=0$ and $n=1$,
which reads
\begin{eqnarray}
\nonumber
%\left \{
 2
{\hat X}^{(tA;0)}_0 +
2{\hat Y}^{(0)}_0+
 {\hat X}^{(tA;1)}_0
%\right \}
&=&
 \left [ L_{+1},  \ {\hat X}^{(tA;0)}_{1}(0) \right ] , \\[1mm]
%\left \{
 2
 {\hat X}^{(tA;1)}_0 + 2 {\hat X}^{(tA;2)}_0
%\right \}
&=&
 \left [ L_{+1},  \ {\hat X}^{(tA;1)}_{1}(0) \right ]  .
\label{RecursionExplicittwithASummarymZeronOne}
\end{eqnarray}
Using the information contained in the lowest order terms of
\rf{tandAXofz} this becomes
\begin{eqnarray}
\nonumber
%\left \{
  h A(0)
%\right \}
&=&
 \left [ L_{+1},  \ {\hat X}^{(tA;0)}_{1}(0) \right ],  \\[1mm]
- 2 h A(0)
%\right \}
&=&
 \left [ L_{+1},  \ {\hat X}^{(tA;1)}_{1}(0) \right ] .
\label{RecursionExplicittwithASummarymZeronOneExplicit}
\end{eqnarray}
The solutions of these equations are
\begin{eqnarray}
\nonumber
{\hat X}^{(tA;0)}_{1}(0)  = - {1\over 2} {\hat X}^{(tA;1)}_{1}(0) &=&
{1\over 2} L_{-1}A(0) + \gamma_1\ {\tilde A}_{h+1}(0)\,,
\\[1mm]
\label{SolutionToOrderZ}
{\hat X}^{(tA;1)}_{1}(0)
\,\,\,\,\,\,
\qquad \qquad
\qquad
 &=& -  L_{-1}A(0) -\delta_1 \ {\tilde B}_{h+1}(0)\,,
\end{eqnarray}
where
$\gamma_1, \delta_1$ are so-far arbitrary coefficients,
and
${\tilde A}_{h+1}(0)$, ${\tilde B}_{h+1}(0)$,
could be any primary operators
of conformal weight $(h+1)$ (if those exist).
Making use of
\rf{RecursionExplicittwithALzeroSummary}
with m=1 shows however that $\delta_1=0$, whereas
$\gamma_1$ remains arbitrary.  ${\tilde A}_{h+1}(0)$
represents the null-vector mentioned above Eq.~\rf{l-L};
using the notation \rf{NotationXandtA},
\be
\label{TildeAhPlusOne}
\ell_{-1} A(0) -{1\over 2} L_{-1}A(0)= \gamma_1\ {\tilde A}_{h+1}(0)\,.
\ee
If the particular theory under consideration
does not have a weight-$(h+1)$ primary operator,
then the extra primary in the first
equation of \rf{SolutionToOrderZ}
is also absent.

\vskip .2cm

\noindent {\it b}) \ \ We continue
with
\rf{RecursionExplicittwithASummary}
for $m=0$ and $n=2$,
which reads
\begin{eqnarray}
\nonumber
\!\!\!\!\!
 4 {\hat X}^{(tA;0)}_0 +
3 {\hat Y}^{(0)}_0+
b A(0)
+ {\hat X}^{(tA;1)}_0
&=&
 \left [ L_{+2},  \, {\hat X}^{(tA;0)}_{2}(0) \right ] , \\[1mm]
 4 {\hat X}^{(tA;1)}_0 + 2 {\hat X}^{(tA;2)}_0
%\right \}
&=&
 \left [ L_{+2},  \, {\hat X}^{(tA;1)}_{2}(0) \right ]  .
\label{RecursionExplicittwithASummarymZeronOneb}
\end{eqnarray}
Using  again the information contained in the lowest-order terms of
\rf{tandAXofz} we arrive at
\begin{eqnarray}
\nonumber
%\left \{
 (2h +b) A(0)
%\right \}
&=&
 \left [ L_{+2},  \, {\hat X}^{(tA;0)}_{2}(0) \right ] , \\[1mm]
 -4h  A(0)
%\right \}
&=&
 \left [ L_{+2},  \, {\hat X}^{(tA;1)}_{2}(0) \right ] .
\label{RecursionExplicittwithASummarymZeronOneExplicitb}
\end{eqnarray}

\vskip .2cm

\noindent {\it c}) \ \ Furthermore,
continuing with
\rf{RecursionExplicittwithASummary}
for $m=1$ and $n=1$,
we get
\begin{eqnarray}
\nonumber
%\left \{
 3 {\hat X}^{(tA;0)}_1 +
2 {\hat Y}^{(0)}_1+
 {\hat X}^{(tA;1)}_1
%\right \}
&=&
 \left [ L_{+1},  \, {\hat X}^{(tA;0)}_{2}(0) \right ] , \\[1mm]
 3 {\hat X}^{(tA;1)}_1 + 2 {\hat X}^{(tA;2)}_1
%\right \}
&=&
 \left [ L_{+1},  \, {\hat X}^{(tA;1)}_{2}(0) \right ]  .
\label{RecursionExplicittwithASummarymZeronOnec}
\end{eqnarray}
Using  the solution \rf{SolutionToOrderZ},
this becomes (upon setting $\alpha=0$)
\begin{eqnarray}
\nonumber
\!\!\!\!\!\!\!\!\!\!\!\!
 3 \left({1\over 2}\right)  L_{-1} A(0)
+ L_{-1}A(0)
= \frac 5 2\,L_{-1}A(0)
&=&
 \left [ L_{+1},  \, {\hat X}^{(tA;0)}_{2}(0) \right ]  ,\\[1mm]
-3 L_{-1}A(0)
%\right \}
&=&
 \left [ L_{+1},  \, {\hat X}^{(tA;1)}_{2}(0) \right ] .
\label{RecursionExplicittwithASummarymZeronOneExplicitc}
\end{eqnarray}

\vskip .2cm

Let us summarize parts b) and c):
the  equations
\rf{RecursionExplicittwithASummarymZeronOneExplicitb}
and \rf{RecursionExplicittwithASummarymZeronOneExplicitc},
\begin{eqnarray}
(2h +b) A(0) &=& \left [ L_{+2},  \, {\hat X}^{(tA;0)}_{2}(0) \right ],
\nonumber\\[1mm]
 -4h  A(0) &=& \left [ L_{+2},  \, {\hat X}^{(tA;1)}_{2}(0) \right ],
 \nonumber\\[1mm]
(5/2)L_{-1}A(0) &=& \left [ L_{+1},  \, {\hat X}^{(tA;0)}_{2}(0) \right ]  ,
 \nonumber\\[1mm]
-3 L_{-1}A(0) &=& \left [ L_{+1},  \, {\hat X}^{(tA;1)}_{2}(0) \right ]  ,
\label{FourUnknownsFourEquations}
\end{eqnarray}
represent four equations for the four unknowns $\alpha^{(0)}, \beta^{(0)}$
and $\alpha^{(1)}, \beta^{(1)}$, which determine
the coefficients  ${\hat X}^{(tA;0)}_{2}(0)$
and ${\hat X}^{(tA;1)}_{2}(0)$, respectively, through
\begin{eqnarray}
\nonumber
{\hat X}^{(tA;0)}_{2}(0)
&=&
\left (
\alpha^{(0)}
L_{-2}
+
\beta^{(0)}
(L_{-1})^2
\right ) A(0)
+
\gamma_2 \ {\tilde A}_{h+2}(0)\,,
\\[1mm]
\label{FourUnknowns}
{\hat X}^{(tA;1)}_{2}(0)
&=&
\left (
\alpha^{(1)}
L_{-2}
+
\beta^{(1)}
(L_{-1})^2
\right ) A(0)\,.
\end{eqnarray}
Here the coefficient $\gamma_2$ remains undetermined, and
${\tilde A}_{h+2}(0)$ is any primary of weight $(h+2)$.
In the notation of \rf{NotationXandtA}
\be
\ell_{-2} A(0) -
\left (
\alpha^{(0)}
L_{-2}
+
\beta^{(0)}
(L_{-1})^2
\right ) A(0)
=
\gamma_2 \ {\tilde A}_{h+2}(0)\,.
\ee
(As before, a similar contribution
to ${\hat X}^{(tA;1)}_{2}(0)$ vanishes by
\rf{RecursionExplicittwithALzeroSummary}.)

\subsection{Details of
the OPE $t(z) t(z) \sim {\bf 1}$}
\label{DetailsofTheOPEtztzero}

The  form of the OPE $t(z) t(z)$ in (\ref{txt})
corresponds to the following coefficients
\begin{eqnarray}
\nonumber
{X}^{(t;0)} &=&
 \quad
 \quad
 \ \
z^2 t(0)
 \qquad
 \qquad
 \ \ \  \
 +
{1\over 2}z^3L_{-1} t(0)
 + O(z^4)\,,
\\[1mm]
\nonumber
{X}^{(t;1)} &=&-2b+ z^2 [-4 t(0)-T(0)]
+
{1\over 2}z^3 L_{-1}[-4 t(0)-T(0)]
 +O(z^4)\,,
\\[1mm]
\label{tandtXofz}
{X}^{(t;2)} &=&
 \quad
 \quad
 \ \
z^2 (-2)T(0)
 \qquad
 \quad
+
{1\over 2}z^3
 (-2)L_{-1}T(0)
+O(z^4)\,.
\end{eqnarray}
Let us derive this OPE from the most singular term,
\be
\label{MostSingulartandt}
{X}^{(t;0)}_0 =0,\ \
{X}^{(t;1)}_0 =-2b,\ \
{X}^{(t;2)}_0 =0\,,
\ee
by applying the recursion (\ref{RecursionExplicitTandTSummary}).

To this end, we first make use of (\ref{LnT}) and (\ref{LnWitht})
with $n=0, 1, 2$
to obtain
\be
[L_{0}, T(0)]=2 T(0),
\qquad
[L_{0}, t(0)]=2 t(0)+ T(0)\,,
\ee
\vskip -8mm
\be
[L_{+1}, T(0)]=0,
\qquad
[L_{+1}, t(0)]=0\,,
\ee
\vskip -8mm
\be
[L_{+2}, T(0)]=0,
\qquad
[L_{+2}, t(0)]=b\,,
\ee
\vskip -8mm
\be
[L_{+2}, L_{-1}T(0)]=0,
\qquad
[L_{+2}, L_{-1}t(0)]=0\,.
\ee
The three equations
(\ref{RecursionExplicitTandTSummary})
read for the special case $m=0, n=2$
\begin{eqnarray}
\nonumber
0 = \left [ L_{+1},  \ {\hat X}^{(t;0)}_{1} \right ]&,& \qquad
\left \{ 0 -2b +3 b \right \}
=
\left [ L_{+2},  \ {\hat X}^{(t;0)}_{2} \right ],
  \\[1mm]
\nonumber
0 = \left [ L_{+1},  \ {\hat X}^{(t;1)}_{1} \right ]&,& \qquad
\left \{ 2 (-2b) + 0 \right \}
=
 \left [ L_{+2},  \ {\hat X}^{(t;1)}_{2} \right ]
,\\[1mm]
\label{RecursionExplicitTandTSummarymiszero}
0 = \left [ L_{+1},  \ {\hat X}^{(t;2)}_{1} \right ]&,& \qquad
\qquad
\qquad
\ \ \
0 = \left [ L_{+2},  \ {\hat X}^{(t;2)}_{2} \right ],
\end{eqnarray}
which
 determines the right-hand side up to primary operators (i.e. $T(0)$ in this case,
 which we added below with undetermined coefficients $\alpha, \beta, \gamma$),
\begin{eqnarray}
\label{solutiontzeroonetwo}
{\hat X}^{(t;0)}_{1} =0&,&
\qquad
{\hat X}^{(t;0)}_{2} = t(0) + \alpha T(0)
\nonumber\,,\\[1mm]
\nonumber
 {\hat X}^{(t;1)}_{1} =0&,&
\qquad
 {\hat X}^{(t;1)}_{2} = -4t(0) + \beta T(0)
\,,\\[1mm]
 {\hat X}^{(t;2)}_{1} = 0&,&
\qquad
 {\hat X}^{(t;2)}_{2} = \gamma  \ T(0)\,.
 \end{eqnarray}

The remaining terms are determined by using
scale invariance, i.e.  by (\ref{RecursionExplicitTandTLzero})
 in the special case $m=2$,
\begin{eqnarray}
\nonumber
2 (t+\alpha T)
+ (-4t + \beta T)
 +
2(2 t + T )
&=&
 \left [ L_{0},  \ {\hat X}^{(0)}_{2}(0) \right ] =
2t +T\,,
\\[1mm]
\nonumber
\!
 2
(-4t + \beta T )
+ 2  \gamma T
&=&
 \left [ L_{0},  \ {\hat X}^{(1)}_{m}(0) \right ] =
(-4)(2t +T) + 2 \beta T
\,,\\[1mm]
\label{RecursionExplicitTandTLzero2}
 2
\gamma T
&=&
 \left [ L_{0},  \ {\hat X}^{(2)}_{m}(0) \right ]
= 2 \gamma T\,,
\end{eqnarray}
or%\\[-8mm]~
\be
 2\alpha+\beta=-1,
\qquad \gamma=-2\,.
\ee
In the first of
(\ref{solutiontzeroonetwo})
we can always change $\alpha$ by redefinitions
as in (\ref{redefine}), and here  we choose $\alpha=0$.
This yields
$\beta =-1,
\gamma=-2
$
in agreement with
(\ref{tandtXofz}).%\\[-8mm]~

\subsection{Subtraction of log and log-squared terms from
$ t(z) t(0)$ OPE}
\label{SectionSubtractionLogs}

In this subsection we consider the contribution of
(the Virasoro representation, or `conformal family'\cite{Belavin1984}
of) the
identity operator to the  linear combination of OPEs
discussed in (\ref{single}) of Section
(\ref{SubSectionTheMultipletOfTwoDimensionalOperatorsAndSusy}),
\be
\label{twithtsubtracted}
t(z) t(0) +4 A_2(z) A_2(0) \ln (z) - T(z)T(0) \ln^2(z) +{1\over 2}  T(z)T(0) \ln(z)\,.
\ee
We have set $\alpha=0$ for convenience (but it can easily
be reinstated).  Our aim is to demonstrate that both single and double powers
of $\ln (z)$ are absent from this expression to all orders in $z$.

\vskip .4cm

$\bullet$ \ \ {\it Log-squared terms:}

It is clear that  the term proportional to
$\ln^2(z)$
cancels;
the expansion coefficients
for this  linear combination are
\be
{\hat X}^{2; total}_m :=
{\hat X}^{(t;2)}_m +4{\hat X}^{(A;1)}_m-{\hat X}^{(T;0)}_m\,.
\ee
The recursion is the {\it same}
for all three summands, so that
we obtain
the
combined recursion
\be
[m+2(n-1)]
{\hat X}^{2; total}_m
=
\left [
L_{+n},
{\hat X}^{2; \,\rm total}_{m+n}
\right ].
\ee
Now, since the expression ${\hat X}^{2; total}_m$
vanishes for small values of $m$, it in fact
%also
vanishes for all values of $m$ as
a consequence of this recursion relation.

%\newpage

\vskip .2cm
\noindent
$\bullet$ \ \ {\it Terms with a single power of log:}

Next consider the terms proportional to
a single power of $\ln(z)$;
the expansion coefficients are
\be
\label{XoneTotal}
{\hat X}^{1; total}_m :=
{\hat X}^{(t;1)}_m
+4
{\hat X}^{(A;0)}_m
+
{1\over 2} {\hat X}^{(T;0)}_m\,.
\ee
The relevant recursion relations are
\begin{eqnarray}
\nonumber
%\left \{
 [m + 2(n-1)] {\hat X}^{(t;1)}_m
+ 2 {\hat X}^{(t;2)}_m
% \right \}
&=&
 \left [ L_{+n},  \ {\hat X}^{(t;1)}_{m+n} \right ],
\\[1mm]
\nonumber
%\left \{
 [m + 2(n-1)]  4 {\hat X}^{(A;0)}_m
%\right \}
+ 4  {\hat X}^{(A;1)}_m
&=&
 \left [ L_{+n},  \ 4 {\hat X}^{(A;0)}_{m+n} \right ],
\\[1mm]
\nonumber
%\left \{
 [m + 2(n-1)] {1\over 2} {\hat X}^{(T;0)}_m
%\right \}
&=&
 \left [ L_{+n},  \ {1\over 2} {\hat X}^{(T;0)}_{m+n} \right ].
\end{eqnarray}
This can be combined into the  {\it key equation}
\be
\label{key}
 [m + 2(n-1)]
 {\hat X}^{1; total}_m
+
 2 \left \{  {\hat X}^{(t;2)}_m + 2 {\hat X}^{(A;1)}_m \right \}
=
 \left [ L_{+n},  \
 {\hat X}^{1; total}_{m+n}
\right ].
\ee
Note that the expression
\be
\label{XttwoAone}
 2 \left \{  {\hat X}^{(t;2)}_m + 2 {\hat X}^{(A;1)}_m \right \}
\ee
is an inhomogeneity, which feeds into the recursion
`externally'.

\vskip .1cm

We see from the third of (\ref{tandtXofz})
and the second of (\ref{AandAXofz}) that
\be
 2 \left \{  {\hat X}^{(t;2)}_m + 2 {\hat X}^{(A;1)}_m \right \}
= z^2 \{ -2T(0)+  2T(0) \} + O(z^3) = 0 + O(z^3).
 \ee
Furthermore,
from the second of
(\ref{RecursionExplicitAtwoAtwoSimplifiedSummary}) and the third
of (\ref{RecursionExplicitTandTSummary}) we see that the quantity
$ \left \{  {\hat X}^{(t;2)}_m + 2 {\hat X}^{(A;1)}_m \right \}$
satisfies the recursion in the second of
(\ref{RecursionExplicitAtwoAtwoSimplifiedSummary}). Since the
$m=0,1,2$ coefficients of the recursion vanish, all coefficients
vanish.  Hence,
 \be
2 \left \{  {\hat X}^{(t;2)}_m + 2 {\hat X}^{(A;1)}_m \right \}
=0\,.
\ee

\vskip .2cm

\noindent {\it In conclusion}, the recursion in \rf{key} now reduces to
\be
 \label{RecursionFinal}
[m + 2(n-1)]
 {\hat X}^{1; \rm total}_m
=
 \left [ L_{+n},  \
 {\hat X}^{1; \rm total}_{m+n}
\right ].
\ee
One finds by inspection that ${\hat X}^{1; total}_m$,
as defined in (\ref{XoneTotal}), vanishes for
small values of the index $m$.
Due to (\ref{RecursionFinal})
this expression then vanishes identically for all values of
the index $m$.\\[-5mm]~

\section*{Appendix C}
%\addcontentsline{toc}{section}{Appendix C}
\renewcommand{\theequation}{C.\arabic{equation}}
\setcounter{equation}{0}

In this appendix we elaborate on the arguments given
below (\ref{Virasorofiveeights}).
Let $\phi(z)$ be a Ka\v{c}-degenerate primary field,
such as e.g. $A_{5/8}(z)$ in (\ref{Virasorofiveeights}).
Let $\Lambda $ be a polynomial in Virasoro lowering operators
$L_{-m}$ ($m\geq 1$),
so that
\be
\bigl ( \Lambda \phi\bigr )(0)
 \ \  {\rm is  \ Virasoro \  primary},
\quad
{\rm i.e} \ \
L_{+n} \bigl (  \Lambda \phi \bigr )(0) =0, \quad (n\geq 1)\,.
\ee
An example is $\Lambda
=$
$ \left(L_{-2}-{2 \over 3} L_{-1}L_{-1}\right)
$
in
Eq.~(\ref{Virasorofiveeights}).

Assume that $\bigl ( \Lambda \phi\bigr )$
can be set to zero when inserted into
any correlation function with other operators.
This implies in particular that
\be
\label{VanishingNullvector}
\langle\bigl ( \Lambda \phi\bigr )(0) \ \phi_1(z_1)\ \phi_2(z_2) \ \phi_3(z_3)
...  \phi_N(z_N)\rangle=0
\ee
for all primary operators $\phi_1,   ...  \phi_N$.
We will show that
\be
\ell_{+n} \bigl (  \Lambda \phi \bigr ) (0) \not =0
\ee
for  {\it some} $n \geq 1$ leads to a contradiction with (\ref{VanishingNullvector}).
To see this, consider the special case where
$ \phi_2 = \phi_1^\dagger $
has non-vanishing two-point function
${<\phi_1(z)\phi_1^\dagger(0)>}\not =0$.
Thus,  we know from (\ref{AA}) and (\ref{VanishingNullvector})
  that
\be
\langle \Lambda \phi(0) \ t(z_2) \ \phi_3(z_3)...  \phi_N(z_N)\rangle=0
\ee
since the insertion of  $T(z)$,
appearing also in the OPE
$ \phi_1(z_1) \phi_1^\dagger(z_2)$,
vanishes due to (\ref{VanishingNullvector}) because $T$ is primary
at central charge $c=0$.
Hence we conclude from the representation (\ref{Defelln})
of $\ell_n$, where the integration contour surrounds
the origin, that
\be
\langle \ell_n \bigl ( \Lambda \phi \bigr )(0)  \ \phi_3(z_3)...  \phi_N(z_N)\rangle=0,
\qquad ({\rm for \ all } \ n \geq 1)
\ee
for all primary operators
$\phi_3,   ...  \phi_N$
(again, due to (\ref{VanishingNullvector})
and because $T(z)$ is primary).  This completes the proof.\\[-8mm]~

\section*{Acknowledgments}

\noindent We  would like to thank Ian Kogan,
John Cardy and Monwhea Jeng for illuminating discussion or
correspondence. This work was supported in part by the NSF through
Grant No. DMR-00-75064 (A.W.W.L).

\newpage

%\addcontentsline{toc}{section}{References}


\begin{thebibliography}{99}

\bibitem{Belavin1984} A. A. Belavin, A. M. Polyakov and A. B. Zamolodchikov, { Nucl. Phys.} B
{\bf 241}, 333 (1984).

\bibitem{DiFrancesco}
P. Di Francesco, P. Mathieu and D. S\'en\'echal, {\it Conformal Field
Theory} (Springer, New York, 1997).

\bibitem{Cardy1986} H. W. J. Bl\"ote, J. L. Cardy and M. P. Nightingale, {Phys. Rev. Lett.}
{\bf 56}, 7462 (1986).

\bibitem{Affleck1986} I. Affleck, {Phys. Rev. Lett.} {\bf 56},
746 (1986).

\bibitem{KnizhnikZamol} V. G. Knizhnik and A. B. Zamolodchikov,
{ Nucl. Phys.} B {\bf  247}, 83 (1984.

\bibitem{1DSpinChains} See e.g.: I. Affleck and F. D. M. Haldane,
{Phys. Rev.} B {\bf 36}, 5291 (1987).


\bibitem{Kondo}  See e.g.: I. Affleck and A. W. W. Ludwig, { Nucl. Phys.} B
{\bf 360}, 641 (1991); A.~W.~W.~Ludwig and I. Affleck { Nucl. Phys.} B
{\bf 428}, 545 (1994).

\bibitem{FreedmanEtAl} M. Freedman, C. Nayak, K. Shtengel, K. Walker and Z. Wang,
{ Ann. Phys.} {\bf 310}, 428 (2004).

\bibitem{EfetovBook} See e.g.:  K. B. Efetov, {\sl Supersymmetry in Disorder and Chaos} (Cambridge
University Press, Cambridge, UK, 1997).

\bibitem{KaneStone} See e.g.:  A. J. McKane and M. Stone, { Ann. Phys.} {\bf 131},
36 (1981).

\bibitem{LeeWangElectronInteractions} D.-H. Lee and Z. Wang,
{ Phys. Rev. Lett.} {\bf 76}, 4014 (1996).

\bibitem{HPWeirecent} R.~T.~F. van Schaijk, A. de Visser, S. Olsthoorn, H.~P.~Wei and A.~M.~M.%
~Pruisken, { Phys. Rev. Lett.} {\bf 84}, 1567 (2000).

\bibitem{HuckesteinRMP} B. Huckestein, { Rev. Mod. Phys.} {\bf 67},
357 (1995).

\bibitem{IQHEnumerical} For a review, see e.g. Ref. \cite{HuckesteinRMP}.

\bibitem{Pruisken} H. Levine, S. B. Libby and A. M. M. Pruisken,
{ Phys. Rev. Lett.} {\bf 51}, 1915 (1983).

\bibitem{Khmelnitskii} D. E. Khmelnitskii, { Zh. Eksp. Teor. Fiz.}
{\bf 38}, 454 (1983); { JETP Lett.} {\bf 38}, 552 (1983).

\bibitem{GruzbergLudwigReadSQHE} I. Gruzberg, A. W. W. Ludwig and
N. Read, { Phys. Rev. Lett.} {\bf 82}, 4524 (1999).

\bibitem{CardySQHE} E. J. Beamond, J. Cardy and J. T. Chalker,
{ Phys. Rev.} B {\bf 65}, 214301 (2002).

\bibitem{ButSee} The conformal spectra of the SUSY theories for
2D percolation and dilute polymers were recently found in:
N. Read and  H. Saleur, {\it Nucl. Phys.} B {\bf 613}, 409 (2001).

\bibitem{ParisiSourlas} G. Parisi and N. Sourlas, { J. Physique Lett.}
(Paris) {\bf  41}, L403 (1980).

\bibitem{Replica} For a review, see e.g.: T. C. Lubensky,
{ in} {\em Les Houches Session XXXI}, eds. Balian, Maynard and Toulouse
(North Holland, Amsterdam 1979).

\bibitem{DeGennes} P. G. deGennes, { Phys. Lett.} A {\bf 38},
339 (1972).

\bibitem{LoopModels} See e.g.: B. Nienhuis, { J. Stat. Phys.}
{\bf 34}, 731 (1984).

\bibitem{SaleurDuplantierLoops} H. Saleur and B. Duplantier,
{ Phys. Rev. Lett.} {\bf 58}, 2325 (1987).

\bibitem{RevModPhysWuPotts} see e.g. F.Y. Wu, { Rev. Mod. Phys.} {\bf 54}, 235 (1982).

\bibitem{qPotts} See e.g.: A. W. W. Ludwig,
{ Nucl. Phys.} B {\bf 330}, 639 (1990);
Vl. S. Dotsenko, M. Pico and P. Pujol, { Nucl. Phys. B}
{\bf 455}, 701 (1995);
J. Cardy and J. L. Jacobsen, { Phys. Rev. Lett.} {\bf 79},
4063 (1997); C. Chatelain and B. Berche, { Nucl. Phys.} B {\bf 572},
626 (2000);  T. Davis and J. Cardy, Nucl. Phys. B {\bf 570} 713 (2000);
M. Jeng and A.~W.~W.~Ludwig, { Nucl. Phys.} B {\bf 594}, (2001).

\bibitem{Gurarie2002} V. Gurarie and A. W. W. Ludwig, { J. Phys. A: Math.
Gen.} {\bf 35}, L377 (2002).

\bibitem{ExplanationOPEBPZ} see e.g. Appendix B of Ref.\cite{Belavin1984},
or  Section \ref{AppBTheOPEofTwoPrimaryOperators} of
Appendix B of this paper.

\bibitem{Gurarie1999} V. Gurarie, { Nucl. Phys.} B {\bf 546},
765 (1999).

\bibitem{Cardy2001} J. L. Cardy, {\em The Stress Tensor in
Quenched Random Systems}, { in} {\em Proceedings~of} {\em the%
~NATO~Advanced Research Workshop on Statistical Field Theories},
Como 2001
(Kluwer, Boston, 2002)\\{}
[http://www-thphys.physics.ox.ac.uk/users/JohnCardy/seminars/como.pdf].

\bibitem{Momo} Monwhea Jeng, private communication.

\bibitem{Gurarie1993} V. Gurarie, { Nucl. Phys.} B {\bf 410},
535 (1993).

\bibitem{Kogan1996} J.-S. Caux, I. I. Kogan and A. M. Tsvelik, {
Nucl. Phys.} B {\bf 466}, 444 (1996).

\bibitem{Kogan2003} I. I. Kogan and A. Nichols, Int. J. Mod. Phys.
A {\bf 18}, 4771 (2003).

\bibitem{Cardy1999} J. L. Cardy, {\em
Logarithmic Correlations in Quenched Random Magnets and Polymers},
cond-mat/9911024.

\bibitem{MonwheaJeng}  {We thank Monwhea Jeng, who
has checked \cite{Momo} these statements up to level $15$, for assistance.}

\bibitem{ZamolodchikovOnIntegrable} A. B. Zamolodchikov,
{ Mod. Phys. Lett.} A {\bf  6}, 1807 (1991).

\bibitem{ZamolodchikovCountingArgument} A. B. Zamolodchikov,
in {\it Advanced Studies in Pure Mathematics} {\bf 19},
K. Aomoto and T. Oda { eds.} (Academic Press, New York, 1989).

\bibitem{ChimZamolodchikov} L. Chim and A. B. Zamolodchikov,
{ Int. J. Mod. Phys.} A {\bf 7}, 5317 (1992).

\bibitem{FendleyRead} P. Fendley and N. Read,
{ J. Phys.} A {\bf 35}, 10675 (2002).

\bibitem{DotsenkoFateev} Vl. S.  Dotsenko, { Nucl. Phys.} B {\bf 235}
(1984) 54; Vl. S.  Dotsenko and V. A. Fateev, { Nucl. Phys.} B {\bf  240}
(1984) 312.

\bibitem{Cardy1991} J. L. Cardy, { J. Phys. A} {\bf 25}, L201
(1992).

\bibitem{Zamolodchikov1986} A. B. Zamolodchikov and A. Fateev,
{ Sov. Phys. JETP} {\bf 62}, 215 (1985) [{Zh. Eksp. Teor.
Fiz.} {\bf 89}, 380 (1985)].

\bibitem{Kogan2002} I. I. Kogan and A. Nichols,
{\em Stress Energy Tensor in $c=0$ Logarithmic Conformal Field Theory},
hep-th/0203207.
\end{thebibliography}
\end{document}